\documentclass[reprint,eqsecnum,floats,aps,amsmath,amssymb%
,footinbib,prd,showpacs]{revtex4-1}

\usepackage{amsmath,amsfonts,amssymb}
\usepackage{enumerate}
\usepackage{calc}
\usepackage{graphicx}
\usepackage{psfrag}

\newcommand{\rd}{{\rm d}}

\newcommand{\integ}{\mathbb{Z}}

\DeclareMathOperator*{\sgn}{sgn}
\newcommand{\inter}{{\rm int}}

\begin{document}

$\hphantom{.}$\vspace{-2cm}
\begin{flushright}
  IGC-11/5-4
\end{flushright}

\title{Effective dynamics of the hybrid quantization of the Gowdy $T^3$ universe}

\author{David \surname{Brizuela}${}^{1}$}
\email{brizuela@gravity.psu.edu}
\author{Guillermo A. \surname{Mena Marug\'an}${}^{2}$}
\email{mena@iem.cfmac.csic.es}
\author{Tomasz \surname{Paw{\l}owski}${}^{3}$}
\email{tpawlows@unb.ca}

\affiliation{
  ${}^{1}$Institute for Gravitation and the Cosmos,
  The Pennsylvania State University,
  104 Davey Lab, University Park, PA 16802, USA
  \\
  ${}^{2}$Instituto de Estructura de la Materia, IEM-CSIC,
  Serrano 121, 28006 Madrid, Spain
  \\
  ${}^{3}$Department of Mathematics and Statistics,
  University of New Brunswick, Fredericton, NB E3B 5A3, Canada
}

\begin{abstract}
  The quantum dynamics of the linearly polarized Gowdy $T^3$ model
  (compact inhomogeneous universes admitting
  linearly polarized gravitational waves) is
  analyzed within Loop Quantum Cosmology by means of an effective
  dynamics. The analysis, performed via analytical and
  numerical methods, proves that the behavior found in the evolution
  of vacuum (homogeneous) Bianchi I universes is preserved qualitatively also in
  the presence of inhomogeneities. More precisely, the initial singularity
  is replaced by a big bounce which joins deterministically two large
  classical universes. In addition, we show that the size of the
  universe at the bounce is at least of the same order of
  magnitude (roughly speaking) as the size of the corresponding homogeneous universe obtained in the
  absence of gravitational waves. In particular, a precise
  lower bound for the ratio of these two sizes is found. Finally,
  the comparison of the amplitudes of the gravitational wave
  modes in the distant future and past shows that, statistically (i.e., for
  large samples of universes), the difference in amplitude is
  enhanced for nearly homogeneous universes, whereas this difference
  vanishes in inhomogeneity dominated cases.
  The presented analysis constitutes the first systematic effective study of an inhomogeneous
  system within Loop Quantum Cosmology, and it proves the robustness
  of the results obtained for homogeneous cosmologies in this context.
\end{abstract}


\maketitle

\section{Introduction}\label{sec:intro}

Loop Quantum
Gravity (LQG) \cite{Rovelli-book,*Thiemann-book,al-status} is one the most
promising approaches for the quantization of gravity. Its development in recent
years provides hopes in the program to overcome the
limits of Classical Relativity, like for example in avoiding the breakdown of physics at
spacetime singularities. In part, this task has already been achieved in the context of the
application of LQG methods to the study of (simple) cosmological systems, known as Loop
Quantum Cosmology (LQC)
\cite{bojo-liv,*a-lqc-overview,*a-lqc-intro,*mm-lqc-overw}. In the
simplest (isotropic) models, the initial big bang singularity has been shown to
be resolved dynamically \cite{aps-prl,aps-det,aps-imp}, being replaced by a big
bounce connecting an expanding universe with a preceding (contracting) one in
a deterministic way. This property, initially described in
flat Friedmann-Robertson-Walker universes with a massless scalar field, was next
confirmed analytically \cite{acs} as well as extended to models of
spherical and hyperbolic topology \cite{apsv-spher,*skl,*van} or with nontrivial
cosmological constant \cite{bp-negL,*kp-posL,*ap-posL}, and even further to
anisotropic homogeneous systems in vacuo \cite{mgp,mgp-ev} or with matter
\cite{s-b1,awe-b1,hp-b1}.

Even within these simple settings, the quantum effects change drastically the standard
understanding of the early universe cosmology, in particular providing solutions
to the horizon problem, conserving the (otherwise violated) entropy bounds \cite{awe-entr},
or increasing the estimate on the probability of inflation as to
become an almost certain event \cite{as-infl}.
It is also expected that the singularity resolution mechanism of LQC extends to
black hole interiors \cite{ab-paradigm,*ab-sch}, extension which may allow
one to cure the long
standing problems of black hole physics (for example preventing potential
information loss) and reveal yet unpredicted phenomena.

In spite of this success, the commented mechanism to avoid singularities
cannot be taken for granted yet since it needs further confirmation in more realistic
situations, which necessarily include inhomogeneities. The few attempts to achieve this
goal that have appeared in the literature
\cite{gb-foot,*mcgb-spectrum,*gcbg-waves,bhks-pert,*bhks-anomaly,*bh-corr}
rely on heuristic constructions, inspired by a polymer quantization, but are not based on
any quantum model that can be considered rigorously defined.

Until recently, the task of building a model
out of a well defined quantum theory, with good understanding and control on
each of the steps of its construction, seemed out of reach, as the most promising
midi-superspace treatments \cite{b-spher,*bs-spher-v,*bs-spher-h} were never
developed past the quantum kinematics and the formal introduction of the
constraints. In the past few years, however, an alternate possibility emerged when a quantization
scheme within the framework of LQC was formulated \cite{mgm-short,mgm,mmw} for a
class of spatially compact cosmological spacetimes admitting linearly polarized
gravitational waves: the $T^3$ Gowdy universes \cite{gowdy-initial}. The
spacetimes of this class admit two spatial Killing fields (with space of orbits
diffeomorphic to $S^1$). Since they are highly symmetric, a symmetry reduction
is allowed. But, even so, they still possess local degrees of
freedom and, furthermore, their structure makes a
nonperturbative analysis viable.

In order to describe these universes quantum mechanically, a so-called {\it hybrid}
quantization scheme was applied (see Ref. \cite{mgm} for details).
First, the geometry was represented as the Fourier modes of gravitational waves
living on a homogeneous (Bianchi I \cite{ch-b1,chv-b1,mgp,awe-b1}) background.
Next, the background geometry was quantized using loop techniques, while a standard
Fock quantization was used to represent the gravitational wave modes \cite{ccmv-gT3}.

Such a quantization prescription allows one to rigorously describe the
system on a kinematical level, and identify from it a Hilbert space of
physical states \cite{mgm} following Dirac's program. However, the complexity of the
quantum constraints and the presence of an infinite number of degrees of freedom,
as well as problems related to the choice of an internal time in vacuo \cite{mgp-ev},
make the analysis of the genuine quantum dynamics extremely difficult.
One of possibilities to deal with these difficulties is to divide the program
into two steps:
\begin{enumerate}[(i)]
  \item A detailed analysis of the dynamical properties of the system by means
    of the so-called \emph{effective dynamics} \cite{sv-eff,*t-eff}, constructed on the basis
    of a \emph{rigorously} defined quantum model. \label{it:step-eff}
  \item The confirmation of the validity of the effective predictions via a \emph{genuine}
    quantum analysis performed in a computationally manageable domain.
\end{enumerate}
The treatment via effective dynamics, while allowing one to adopt the procedures and methods of
classical mechanics, incorporates the quantum effects of the
discrete geometry to a certain extent. In the cases tested so far, this kind of effective dynamics
has proven to reproduce the genuine quantum dynamics of states with a
semiclassical behavior up to a remarkable accuracy
\cite{aps-imp,acs,bp-negL,s-b1,mgp-ev}. On the other hand, the results of
this effective approach, even when regarded as preliminary, make accessible and unveil
properties of the system which are important for the correct formulation and
refinement of the genuine theory.

In this article, we focus on step \eqref{it:step-eff}. Starting from the quantum model obtained
with the hybrid quantization, we construct the effective, classical dynamical system which
approximates the quantum dynamics to first order (i.e. without taking into account
the effect
of state-dependent parameters on the dynamics). We then apply the resulting description to address
the following questions. Are the cosmological singularities dynamically resolved by a big bounce
mechanism in the presence of the (inhomogeneous) gravitational waves? And,
if the answer is in the affirmative, does this bounce happen under conditions
which are similar to those found in homogenous scenarios? Or do there exist substantial
modifications? Furthermore, if the bounce persists, how does it affect
the gravitational wave modes? In particular, we are going to focus our attention
on the problem of whether the energy distribution of the modes
change significantly through the bounce process.

The main conclusions of this study were briefly reported in Ref.~\cite{bmp-letter}.
In this article, we present the details of the analysis and the methodology employed in it.
This material, while important to validate the results of Ref. \cite{bmp-letter},
also provides tools which are useful in the analysis of more complicated
(and more realistic) inhomogeneous cosmological systems,
e.g. the corrected vacuum Gowdy universe description of Ref. \cite{mmw},
as well as its extension incorporating matter fields \cite{mmm}.
The results themselves constitute only a qualitative guideline to
elucidate the kind of phenomena that are observed in the considered
class of inhomogeneous systems. The fine details of these results
must not be regarded as providing definitive answers,
in particular owing to the specific quantization prescription that
has been adopted (see the discussion in Sec.~\ref{sec:frame-hybrid}).
The system obtained with the corrected quantization prescription of
Ref. \cite{mmw}, however, has a sufficiently similar
structure \cite{mmw} and behavior \cite{hp-b1} as to expect that our
results are still applicable to it, at least qualitatively.

The paper is organized as follows. We start by providing a
brief introduction to the hybrid quantization of the
model in Sec.~\ref{sec:frame}. Next, in Sec.~\ref{sec:eff} we
introduce its effective description, deriving
and discussing the structure of the effective equations of motion (EOM)
and their asymptotic behavior. In order to provide a reference
for comparison, these equations are analytically solved in
the absence of inhomogeneities in Sec.~\ref{sec:B1},
thus integrating the effective dynamics of the Bianchi I homogeneous background.
The core of the article is presented in two sections.
In Sec.~\ref{sec:bounce} we combine analytical and numerical
methods to discuss how the presence of inhomogeneities affects
the appearance and properties of the bounce. On the other hand,
Sec.~\ref{sec:dyn} is devoted to numerical studies of the
dynamical trajectories. Finally, in Sec.~\ref{sec:inh} we study the effect of
the bounce on the structure of the inhomogeneities by Monte-Carlo methods.
In Sec.~\ref{sec:concl}, we provide a summary of the
results, as well as a discussion of their
relevance, limitations, and possible extensions.

\section{Hybrid Gowdy model}
\label{sec:frame}

In this section we briefly introduce the class of Gowdy universes of
$T^3$ topology quantized within the LQC framework. We start by
specifying the model at the classical level in Subsec.~\ref{sec:frame-class},
and outline then the main steps and properties of the hybrid quantization
in Subsec.~\ref{sec:frame-hybrid}.

\subsection{The Gowdy $T^3$ universe}\label{sec:frame-class}

The general Gowdy models are
spacetimes with spatial sections of
compact topology and two spatial Killing vector fields. In this article, we restrict
our attention just to one model: the vacuum model with the spatial
topology of a $3$-torus ($T^3$).
Besides, we will consider only the subfamily of universes whose
content of gravitational waves is linearly polarized. These restrictions imply
that the Killing fields are axial and hypersurface orthogonal.

Within this family of spacetimes, we can make use of a distinguished coordinate
system $(t,\theta,\sigma,\delta)$, where $\sigma$ and $\delta$ are axial coordinates adapted
to the Killing fields. In a $3+1$ decomposition, owing to the hypersurface orthogonality,
the induced $3$-metric can be chosen to be diagonal, and one characterizes it just by three fields
corresponding to the norm of one of the Killing fields, the area of the isometry group orbits,
and the scale factor of the $1$-metric induced on the manifold of group orbits. Most of the gauge
freedom is fixed by the condition that both the generator of conformal transformations of the considered
$1$-metric as well as the area of the orbits be homogeneous functions. In fact this leaves out only
two gauge degrees of freedom, namely, those corresponding to
the spatial averages of the Hamiltonian constraint and of the
diffeomorphism acting in the $\theta$ direction.

The phase space of the system splits into two sectors formed, respectively,
by the following degrees of freedom
\begin{enumerate}[a)]
  \item \textbf{Homogeneous:}
    the spatial averages (zero Fourier modes) of the metric fields and their conjugate variables.
  \item \textbf{Inhomogeneous:} all the nonzero modes of the only metric field
    $\xi(\theta)$ whose spatial dependence has not been determined in the gauge-fixing
    procedure (namely, the norm of one of the Killing vector fields),
    together with the conjugate momenta of these modes.
\end{enumerate}
The homogeneous sector describes the phase space of a vacuum Bianchi I model with
$3$-torus topology \cite{mm-ijmpa}, to which the considered Gowdy model reduces when
all the inhomogeneities vanish. In order to represent it,
we employ the standard LQC description of the Bianchi I model \cite{mgp}, expressing
the degrees of freedom in terms of Ashtekar-Barbero variables. These are
an $SU(2)$ valued connection $A^a_i$, and a densitized triad $E^i_a$.
Upon fixing the gauge by the requirement that the $3$-metric be diagonal and
with the introduction of a fiducial Euclidean metric, these variables get determined by (time varying)
connection and triad coefficients:
\begin{equation}
  A^a_i = \frac{c^i}{2\pi}\delta^a_i , \quad
  E^i_a = \frac{p_i}{4\pi^2}\delta^i_a .
\end{equation}
Here, $a$ is an internal $SU(2)$ index, $i=\theta, \sigma, \delta$, and no summation is assumed
for repeated indices. These coefficients
are canonically conjugate, with Poisson brackets $\{c^i,p_j\}=8\pi G\gamma\delta^i_j$, where
$\gamma$ is the Immirzi parameter.

The degrees of freedom of the inhomogeneous sector are encoded in the metric field
$\xi(\theta)$ and its conjugate momentum $P_{\xi}(\theta)$, both with the zero mode excluded.
They can be decomposed into (nonzero) Fourier modes, obtaining an infinite countable set of
canonical pairs, $\{(\xi_m,P_{\xi}^m),m\in\integ\setminus\{0\}\}$.
For these modes, we introduce creation and annihilation variables $(a_m,a^*_m)$ in analogy to those
naturally associated with a massless scalar field:
\begin{equation}
  a_m = \sqrt{\frac{\pi}{8G|m|}} \left( |m|\xi_m + i \frac{4G}{\pi} P_{\xi}^m \right) .
\end{equation}
It is easy to check that they have the Poisson brackets
$\{a_m^{},a^*_{\tilde{m}}\}=-i\delta_{m\tilde{m}}$.

With this choice of variables, the spatial metric takes the form:
\begin{equation}\begin{split}
  \rd s^2 = \frac{|p_\theta p_\sigma p_\delta|}{4\pi^2}
  &\left[ e^{\tilde{\gamma}}
    \left(-\frac{N_{_{_{\!\!\!\!\!\!\sim}}\;}^2}{(2\pi)^4}\rd t^2 + \frac{\rd\theta^2}{p_\theta^2} \right)
  \right. \\
  &\left.\hphantom{=} +\, e^{-\frac{2\pi\tilde{\xi}}{\sqrt{p_{\theta}}}} \frac{\rd\sigma^2}{p^2_{\sigma}}
  + e^{\frac{2\pi\tilde{\xi}}{\sqrt{p_{\theta}}}} \frac{\rd\delta^2}{p^2_{\delta}} \right] ,
\end{split}\end{equation}
where $N_{_{_{\!\!\!\!\!\!\sim}}\;}$ is the densitized lapse function,
\begin{equation}
  \tilde{\xi}(\theta) = \sum_{m\neq 0} \frac{\sqrt{G}}{\pi\sqrt{|m|}}
  (a_m + a^*_{-m} ) e^{im\theta}
\end{equation}
is the contribution of the nonzero modes to the field $\xi(\theta)$,
and the function $\tilde{\gamma}(\theta)$ equals
\begin{subequations}\begin{align}
  \begin{split}
    \tilde{\gamma}(\theta) &= \left(\frac{2c_{\delta}p_{\delta}}{c_{\sigma}p_{\sigma}+c_{\delta}p_{\delta}}
      -1 \right) \frac{2\pi}{\sqrt{|p_{\theta}|}} \tilde{\xi}(\theta) \\
    &\hphantom{=} -\frac{\pi^2}{|p_{\theta}|} \tilde{\xi}^2(\theta)
    - \frac{8\pi G\gamma}{c_{\sigma}p_{\sigma}+c_{\delta}p_{\delta}} \zeta(\theta) ,
  \end{split} \\
  \begin{split}
    \zeta(\theta) &= i \sum_{m,\tilde{m}\neq 0} \sgn(m+\tilde{m})
      \frac{\sqrt{|m+\tilde{m}||\tilde{m}|}}{m} \\
    &\hphantom{=} \times
    (a_{-\tilde{m}}-a_{\tilde{m}}^*)(a_{m+\tilde{m}}+a_{-(m+\tilde{m})}^*)
    e^{im\theta} .
  \end{split}
\end{align}\end{subequations}

The only nontrivial constraints of the system --the spatial averages
of the densitized Hamiltonian constraint $\mathcal{C}_G$ and of the
diffeomorphism constraint in the $\theta$ direction $C_{\theta}$
(corresponding to rigid rotations in $\theta$)-- can be written
\begin{subequations}\label{momentum}\begin{align}
  C_{\theta} &= \sum_{m=1}^{\infty} m (a_m^* a_m^{} - a_{-m}^* a_{-m}^{})  ,
  \label{eq:diff-constr} \\
  \begin{split}
    \mathcal{C}_G
    &= -\frac{1}{\gamma^2} \sum_{i\neq j} c^i p_i c^j p_j  \\
    &\hphantom{=} + G\left[
      \frac{c^\sigma p_\sigma+c^\delta p_\delta}{\gamma^2|p_{\theta}|} H^{\xi}_{\inter}
      +32\pi^2|p_{\theta}|H^{\xi}_0
      \right] .
  \end{split} \label{eq:ham-constr}
\end{align}\end{subequations}
In these equations, all the information about the inhomogeneous degrees of freedom is contained
in the terms
\begin{subequations}\label{eq:Hdefs}\begin{align}
  \label{H0def}
    H_0^\xi &\equiv  \sum_{m=1}^{\infty} m \left[a_m^*a_m^{} + a_{-m}^*a_{-m}^{}\right] , \\
  \label{Hintdef}
  \begin{split}
    H_{\rm int}^\xi &\equiv \sum_{m=1}^{\infty}\frac{1}{m}
      \left[ a_m^*a_m^{}+a_{-m}^*a_{-m}^{}+a_{m}^*a_{-m}^*
      \right.   \\
    &\hspace{1.6cm}+  \left. a_{m}^{}a_{-m}^{}\right]  ,
  \end{split}
\end{align}\end{subequations}
which have the form of a free field Hamiltonian and of an interaction Hamiltonian (which creates and
annihilates pairs of particles). Finally, defining $\Theta_i\equiv c_i p_i$, we can express the
classical Hamiltonian constraint $\mathcal{C}_G$ in a more compact form:
\begin{equation}\label{Hamiltonian}\begin{split}
  {\cal C}_G &= -\frac{2}{\gamma^2}(\Theta_\theta \Theta_\delta+
    \Theta_\theta \Theta_\sigma+\Theta_\delta \Theta_\sigma)  \\
  &\hphantom{=}
    + \frac{G}{\gamma^2} (\Theta_\delta + \Theta_\sigma)^2 \frac{1}{|p_\theta|} H_{\rm int}^\xi
    +32 \pi^2 G |p_\theta| H_0^\xi  .
\end{split}\end{equation}

At this point, it is worth commenting that, if one adopts the same choice of lapse as in full
LQG (namely $N=1$), the zero mode that one obtains for the Hamiltonian constraint
is $\mathcal{C}_G/V$, rather than $\mathcal{C}_G$ (with $V$ being the Bianchi I volume
$V:=\sqrt{|p_\theta p_\delta p_\sigma|}$). In fact, the constraint $\mathcal{C}_G/V$
is indeed the one which is implemented quantum mechanically in the treatment applied
in Refs. \cite{aps-imp,mgp}. The counterpart of $\mathcal{C}_G$ is then constructed by
introducing a scaling by $V$ at the quantum level.

\subsection{Hybrid quantization}\label{sec:frame-hybrid}

The quantization of this Gowdy model was achieved recently adopting a hybrid approach
\cite{mgm-short,mm-ijmpa,mgm,mmw}, which is based on the hypothesis that the most relevant quantum
geometry effects are those affecting the homogeneous sector representing the Bianchi I
background on which the gravitational waves propagate. In this way, the zero modes of the geometry
are quantized using the techniques of LQG, while a standard Fock quantization is
employed for the inhomogeneities arising from the presence of gravitational waves.

The proposed quantum model is thus applicable in the regime where the ``energy'' of the inhomogeneities is
distributed among the modes so that, while the overall backreaction of the background may feel the discreteness
effects, the energy of each individual mode is still small for the polymer effects on it
to become important.
Remarkably, this hybrid quantization proves to be viable and consistent. The kinematical Hilbert space
of the system is constructed as the product of the kinematical Hilbert space for Bianchi I
(with $T^3$-topology) in LQC and the Fock space for the inhomogeneous sector. The two sectors are
coupled by the Hamiltonian constraint, so that the quantization is indeed nontrivial.

\subsubsection{The Bianchi I background}

Let us first describe the polymeric quantization of the Bianchi I geometry, i.e., of
the homogeneous sector. In this quantization scheme, the elementary variables are
fluxes, which are proportional to the triad coefficients $p_i$, and holonomies of connections
computed along edges of coordinate length $2\pi\mu_i$ in each of the (gauge-fixed)
directions $i$, where $\mu_i$ is any real number. The elements of these holonomies
are linear combinations of the exponentials
$\mathcal N_{\mu_i}(c_i)=\exp(i\mu_{i}c_{i}/2)$
(no Einstein summation convention is adopted). In the triad representation, each of
these elements is represented by an eigenstate $|\mu_{i}\rangle$.
The completion of the vector space spanned by these states with respect to
the discrete inner product $\langle\mu_i|\mu_i^\prime\rangle=\delta_{\mu_i \mu_i^\prime}$ provides
the kinematical Hilbert space, $\mathcal H_{\text{Kin},i}$, for
each direction. The kinematical Hilbert space of the homogeneous sector is simply
$\mathcal{H}_{\text{Kin}}=\otimes_i \mathcal H_{\text{Kin},i}$.

On the basis of states $|\mu_i\rangle$, the action of the
elementary operators is
\begin{equation}\label{action}
  \hat p_i|\mu_i\rangle=4\pi\gamma
  l_\text{Pl}^2\mu_i|\mu_i\rangle,
  \qquad
  \hat{\mathcal N}_{\mu_i^\prime}|\mu_i\rangle=|\mu_i
  +\mu_i^\prime\rangle.
\end{equation}
Here, $l_\text{Pl}=\sqrt{G\hbar}$ is the Planck length. The Hamiltonian constraint is constructed by
defining curvatures in terms of holonomies along closed loops, formed by straight edges in the different
directions $\theta$, $\sigma$, and $\delta$, and regularizing the inverse volume by means of commutators of
holonomies with the volume operator \cite{abl-lqc,al-status,ch-b1}.

To encode the discreteness of the geometry characteristic of LQG, in the above
construction of the curvature operator in terms of holonomies, the area enclosed by the loops is assumed
to coincide with the minimum nonzero eigenvalue $\Delta$ of the area operator in LQG. This
prescription is usually called the {\it improved dynamics} approach.
As a consequence, the coordinate length of
the holonomy along each edge takes a value $2\pi\bar\mu_i$ which depends on the considered state, because
so does the area.

In the process of arriving to a proper understanding of the anisotropic models in LQC, different ways
to implement this prescription have been presented in the literature \cite{ch-b1,awe-b1}. We will adopt
here the proposal explored, e.g., in Refs. \cite{ch-b1,ch-b1eff,mgp},
according to which
\begin{equation}\label{mubarra}
  {\frac1{\bar\mu_i}}=\frac{\sqrt{|p_i|}}{\sqrt{\Delta}}.
\end{equation}
Although this proposal has proven to lead to physical inconsistencies in the case of
noncompact spatial sections (see e.g. \cite{cs-uniq-b1}), we will follow it here owing to
two reasons: $(i)$ its simple mathematical structure provides a well controllable way of combining the
resulting background with the Fock theory which describes the inhomogeneities, and $(ii)$ the proposal
that corrects its drawbacks, put forward in Ref.~\cite{awe-b1},
is similar enough in structure \cite{hp-b1} as to expect that the
qualitative results obtained here will also hold for it.
We also note that the commented problems of inconsistency are not present in our system, owing to
the $T^3$-topology \footnote{It should be clear that the effects of the
compact topology by themselves do not justify a priori adopting this quantization prescription,
but rather the similarity between its relevant features
and those of the prescription of Ref. \cite{awe-b1}.},
and that the possible discrepancies with the proposal of Ref. \cite{awe-b1}
should disappear in the region of large triad variables.

Given that $\bar\mu_i$ depends on the triad coefficients, the
operator $\hat{\mathcal N}_{\bar\mu_i}$ generates in fact a
state dependent transformation on the basis formed by
$|\mu_\theta,\mu_\sigma, \mu_\delta\rangle=\otimes_i|\mu_i\rangle$.
Fortunately, the generators (in phase space) of the corresponding shifts
commute, so that one can relabel this basis using affine parameters $\textrm{v}_i$ instead of
the labels $\mu_i$, in such a way that the action of
$\hat{\mathcal N}_{\bar\mu_i}$ is given just by a constant displacement
of the new label $\textrm{v}_i$.
These affine parameters are related to the $p_i$'s as follows
\begin{equation}\label{eq:v-def}
  \textrm{v}_i := \sqrt{\frac{16\pi\gamma l_{\text{Pl}}^2}{9\Delta}}
  \, \text{sgn}(p_i)|p_i|^{3/2}.
\end{equation}
With this relabeling, the action of the basic operators becomes \cite{ch-b1,mgp}
\begin{subequations}\label{eq:bas-op-v}\begin{align}
  \hat{\mathcal N}_{\pm\bar\mu_i}|\textrm{v}_i\rangle&=|\textrm{v}_i\pm1\rangle,\\
  \label{repA}
  \hat p_i|\textrm{v}_i\rangle&=(6\pi\gamma l_{\text{Pl}}^2\sqrt{\Delta})^\frac{2}{3}
  \text{sgn}(\textrm{v}_i)|\textrm{v}_i|^{\frac2{3}}|\textrm{v}_i\rangle.
\end{align}\end{subequations}
On the other hand, the regularized triad operator, representing the inverse of $p_i$,
takes the form:
\begin{equation}\label{inv}
  \widehat{\left[\frac{1}{\sqrt{|p_i|}}\right]}
  |\textrm{v}_i\rangle =b(\textrm{v}_i)
  |\textrm{v}_i\rangle,
\end{equation}
where
\begin{equation}
  b(\textrm{v}_i)=\frac{3|\textrm{v}_i|^{1/3}}{2({6\pi \gamma l_\text{Pl}^2\sqrt{\Delta}})^{1/3}}
  \left||\textrm{v}_i+1|^{1/3}-|\textrm{v}_i-1|^{1/3}\right|.
\end{equation}

It is possible to see \cite{mm-ijmpa,mgm} that, with a suitable choice of symmetric factor ordering,
the Hamiltonian constraint attained for the Bianchi I model with this quantization approach (and
with the lapse chosen equal to the unity) leaves
invariant the superselection sectors formed by the states which have support only on (the product
over the directions $i$ of) semilattices of the form
\begin{equation}
\mathcal
L_{\varepsilon_i}^\pm=\{\pm(\varepsilon_i+2k),
k=0,1,2...\},
\end{equation}
where $\varepsilon_i$ can be any real number in the interval $(0,2]$. Note that the
states with a vanishing $\textrm{v}_i$ decouple from these superselection sectors.
In particular, this allows one to introduce the inverse of the operator \eqref{inv} in each of those
sectors. In turn, this enables one to make a change of densitization in the Hamiltonian constraint
\cite{mgm-short,mm-ijmpa,mgm} which provides
us, finally, with a quantum counterpart of the Bianchi I part of the constraint $\mathcal{C}_G$,
introduced in Eq.~\eqref{eq:ham-constr}. This constraint is obtained by representing the classical
quantities $\Theta_i$ by the operators:
\begin{equation} \label{thetaquan}
  \widehat{\Theta}_i=-\frac{i}{4\sqrt{\Delta}}
  \widehat{\left[\frac{1}{\sqrt{|p_i|}}
  \right]}^{-\frac1{2}}\widehat{\sqrt{|p_i|}}\widehat{M}_i
  \widehat{\sqrt{|p_i|}}\widehat{\left[\frac1{\sqrt{|p_i|}}
  \right]}^{-\frac1{2}}
\end{equation}
with
\begin{equation}\begin{split}
  \widehat{M}_i &= \bigg[(\hat{\mathcal{N}}_{2\bar\mu_i}-\hat{\mathcal{N}}_{-2\bar\mu_i})
    \widehat{\text{sgn}(p_i)}  \\
  &\hspace{0.7cm} +\widehat{\text{sgn}(p_i)}(\hat{\mathcal{N}}_{2\bar\mu_i}-\hat{\mathcal{N}}_{-2\bar\mu_i})\bigg] .
\end{split}\end{equation}
Note that all these operators commute among themselves, and $\widehat{\Theta}_i$ commutes with $p_j$ for $i\neq j$.

\subsubsection{The inhomogeneities}

We next proceed to introduce and quantize the inhomogeneities of the system. This is achieved by constructing
a symmetric Fock space $\mathcal F$, where the
creation and annihilation variables $\{(a_m,a_m^*)\}$ for all the nonzero modes are promoted
to creation and annihilation operators acting on the vacuum.
This Fock quantization is the only one (up to equivalence)
which respects the invariance of the vacuum under rigid rotations
in $\theta$ and ensures that (after deparametrization) the field dynamics is unitarily implemented
\cite{ccmv-gT3-uniq,*cmv-gT3-uniq}.

The $n$-particle states $|\{n_m\}\rangle:=|...,n_{-m},...,n_m,...\rangle$ provide an orthonormal
basis for this Fock space. Here $n_m<\infty$ is the
occupation number of the $m$-th mode, and it is assumed that
only a finite set of these occupation numbers is different from zero.

Adopting normal ordering, the constraint $C_{\theta}$ given in Eq.~\eqref{eq:diff-constr},
which is the generator of rigid rotations, becomes the quantum operator
\begin{equation}\label{quantumCtheta}
  \widehat C_\theta=\sum_{m=1}^\infty m (\hat
  a^\dagger_m \hat a^{}_m-\hat a^\dagger_{-m} \hat a^{}_{-m}).
\end{equation}
The kernel of this operator is a proper Fock subspace.

In the total kinematical Hilbert space $\mathcal{H}_{\text{Kin}} \otimes \mathcal{F}$, we can follow
the quantization procedure
explained above and construct the quantum constraint representing \eqref{eq:ham-constr}. This constraint
is obtained by adopting normal ordering in the operator representation of the inhomogeneous
contributions \eqref{H0def} and \eqref{Hintdef},
and by substituting $p_\theta$ and $\Theta_i$ by their quantum counterparts \eqref{repA} and \eqref{thetaquan}.

Finally, the physical Hilbert space of the system can be found, e.g., by realizing that one can identify
solutions of the Hamiltonian constraint with initial data on the section of minimum value of $\textrm{v}_\theta$,
namely $\textrm{v}_{\theta}=\varepsilon_{\theta}$. The Hilbert structure in the space of initial data can be determined
by imposing self-adjointness conditions on physical observables \cite{mm-ijmpa,mgm}.
Then, on the constructed Hilbert space, one can easily impose the remaining constraint $C_{\theta}$,
completing the quantization of the model.

\section{Effective Gowdy dynamics}
\label{sec:eff}

In this section we introduce the effective EOM
that result from the hybrid quantization of the Gowdy spacetimes, and solve them
in certain regimes of interest. This effective
treatment is going to be classical, in the sense that we are going to define an
effective classical (polymerized) Hamiltonian and the evolution of the relevant quantities
is going to be provided by Hamilton-Jacobi equations. Even so,
this Hamiltonian captures in fact relevant information about the modifications to General Relativity that
are expected to mimic the (loop) quantum behavior of the system.

\subsection{Equations of motion}
\label{sec:eoms}

Given the quantum treatment of the system presented in Sec.~\ref{sec:frame-hybrid}, we
now proceed with the construction of its effective description. We ignore all the
state-dependent quantities, thus building only a first-order description. This amounts
to replacing the basic operators, that is $\hat{p}_i$ and $\mathcal{N}_{\pm\bar{\mu}_i}$
[see Eq. \eqref{eq:bas-op-v}], as well as $\hat{a}_m^{}$ and $\hat{a}_m^{\dagger}$,
by their expectation values. This constitutes
an approximation which should be valid for very sharply peaked states. To simplify the notation,
we will further drop the brackets $\langle\cdot\rangle$ in the expectation values,
denoting for example $\langle \hat{p}_i \rangle$
by $p_i$, and so on.

Furthermore, since we are interesting in solutions corresponding to macroscopic universes, we focus our attention on the regime
$\textrm{v}_i\gg 1$, thus dropping the corrections resulting from the regularization \eqref{inv}.
This procedure leads to a set of constraints of the same form as in Eqs. \eqref{eq:diff-constr},
\eqref{eq:Hdefs}, and \eqref{Hamiltonian}
where the terms $\Theta_i$ (that in the classical theory are equal to $c_i p_i$)
take the polymerized form
\begin{equation}\label{eq:Theta-eff}
  \Theta_i \equiv p_i\frac{\sin(\bar\mu_i c_i)}{\bar\mu_i},
\end{equation}
where $\bar\mu_i = M/\sqrt{|p_i|}$ with $M = \sqrt{\Delta}$, according to Eq. \eqref{mubarra}.

At this stage, it is worth recalling that our effective treatment encapsules the properties of the quantum
description from which it originates. It is thus based on the prescription of Refs. \cite{ch-b1,chv-b1} used to determine
$\bar{\mu}_i$. One can however apply directly the techniques used here to the corrected prescription
put forward in Ref. \cite{awe-b1}, resulting in an effective model whose relevant qualitative features (for our study)
are in fact similar.

Let us now  rewrite the Hamiltonian constraint in terms of the canonical pair
\begin{equation}\label{eq:new-vb}
  b_i = \frac{M c_i}{\sqrt{|p_i|}},\qquad v_i = \frac{2}{3 M} p_i\sqrt{|p_i|},
\end{equation}
where $\{b_i,v_j\}=8\pi\gamma G\delta_{ij}$ and $v_i$ is related with the
affine parameter $\textrm{v}_i$ defined in Eq. \eqref{eq:v-def} via a rescaling by a constant.
It is then straightforward to see that $\Theta_\delta$ and $\Theta_\sigma$
are the only terms in the Hamiltonian constraint that involve $(b_\delta,v_\delta)$ and $(b_\sigma,v_\sigma)$,
respectively; thus they are constants of motion.
For the rest of variables, we obtain the effective EOM just by
taking their Poisson brackets with the Hamiltonian constraint:
\begin{widetext}
\begin{subequations}\label{eq:EOMs}\begin{align}
  \label{vthetaeq}
  \dot{v}_\theta &= \frac{24\pi G}{\gamma}
  \left(\Theta _{\delta}+\Theta _{\sigma}\right) v_\theta\cos b_\theta,
  \\
  \label{bthetaeq}
  \begin{split}
    \dot{b}_\theta &= -\frac{24 \pi G}{\gamma}
    \left(\Theta _{\delta}+\Theta _{\sigma }\right)\sin b_\theta
    -\frac{16\pi G^2}{3\gamma}\left(\frac{2}{3M}\right)^{2/3}
    \left(\Theta_{\delta }+\Theta _{\sigma }\right)^2
    \frac{{\rm sgn} (v_\theta)}{|v_\theta|^{5/3}}H_{\rm int}^\xi
    \\&\hphantom{=}
    + \frac{512}{3} \pi^3 G^2\gamma\left(\frac{3 M}{2}\right)^{2/3}
    \frac{{\rm sgn} (v_\theta)}{|v_\theta|^{1/3}} H_0^\xi,
  \end{split}
  \\
  \label{aeq}
  \dot{a}_m^{} &= -32\pi^2 i G |m|
  \left(\frac{3 M}{2}\right)^{2/3} v_\theta^{2/3} a_m^{}
  -\frac{i G}{\gamma^2|m| v_\theta^{2/3}}\left(\frac{2}{3M}\right)^{2/3}
  \left(\Theta_{\delta }+\Theta _{\sigma }\right)^2(a_m^{}+a_{-m}^*),
  \\
  \label{vdeltaeq}
  \dot{v}_\delta &= \frac{24 \pi G}{\gamma}
  \left(\Theta _{\theta }+\Theta _{\sigma}-\left(\frac{2}{3M}\right)^{2/3} G
  \left(\Theta _{\delta }+\Theta _{\sigma}\right)
  \frac{H_{\rm int}^\xi} {v_\theta^{2/3}} \right)v_\delta \cos b_\delta,
  \\
  \label{bdeltaeq}
  \dot{b}_\delta &= -\frac{24 \pi G}{\gamma}
  \left(\Theta _{\theta }+\Theta _{\sigma}-\left(\frac{2}{3M}\right)^{2/3} G
  \left(\Theta _{\delta }+\Theta _{\sigma}\right)
  \frac{H_{\rm int}^\xi}{v_\theta^{2/3}}\right)\sin b_\delta.
\end{align}\end{subequations}
\end{widetext}
The evolution equations for the conjugate pair $(v_\sigma, b_\sigma)$ are equivalent to those of
$(v_\delta, b_\delta)$; one only has to make the replacements $(\delta\rightarrow\sigma, \sigma\rightarrow\delta)$.
The equation for $a_{m}^*$ is given by the complex conjugate of Eq. \eqref{aeq}.
In order to recover the classical (noneffective) EOM, it suffices to realize
that the dynamics dictated by the original classical constraint \eqref{Hamiltonian}
is given by Eqs. \eqref{eq:EOMs} after replacing
$\sin b_i$ with $b_i$ and $\cos b_i$ with 1 (this can easily be checked by inspection).
Remarkably, the form of the
EOM for the inhomogeneities $a_m$ is not changed
by the effective dynamics for each given value of the pair of constants of motion
$\Theta_\delta$ and $\Theta_\sigma$, i.e., their form is exactly the same as in the genuine classical case.

At this point, it is helpful to inspect the coupling between the different
equations. The EOM for the inhomogeneities and the variables $\{v_\theta,b_\theta\}$ form
a system that is decoupled from the rest, since the variables in the homogeneous directions
$\delta$ and $\sigma$ only enter these equations through $\Theta_\delta$ and $\Theta_\sigma$.
On the other hand, the equations for the variables
corresponding to the coordinates $\delta$ and $\sigma$ are decoupled from each other, but coupled to
the first system of equations through $H_0^\xi$, $H_{\rm int}^\xi$, and $v_\theta$.
In Sec.~\ref{sec:inh} we will make use of this fact to solve the system defined
by $\{v_\theta,b_\theta, a_m\}$ by their own, and analyze in this way
the behavior of the inhomogeneities in a passage through the bounce.

The system also admits another set of constants of motion \cite{mgm}, namely,
\begin{equation}\label{km}
  K_m\equiv a_m^{} a_m^*-a_{-m}^{} a_{-m}^*.
\end{equation}
Quantum mechanically, each of them represents the difference
between the number of particles in the modes $m$ and $-m$.
The conservation of these quantities follows from the decoupling of the modes with different value
of $|m|$ in the dynamics and the conservation of the total field momentum,
as required by the constraint
\begin{equation}\label{constraint}
  {C}_{\theta} \equiv \sum_{m=1}^\infty m K_m .
\end{equation}

\subsection{Asymptotic regime}\label{sec:asympt}

One of the purposes of our analysis is to compare the properties of the universe in the distant future and past,
regions where its evolution is expected (or at least hoped) to follow the standard classical trajectory.
Therefore, with the aim at verifying this expectation, as well as to build suitable methods for the comparison
at such distant epochs, it is useful to study
the ``low-curvature'' limit of the EOM. This limit corresponds to
small values of $b_\theta$ (for which the standard classical EOM are reproduced)
and large $v_\theta$ (providing large volumes). In this limit, it is straightforward to solve Eq. \eqref{vthetaeq},
\begin{equation}
  v_\theta = v_\theta(0)\exp\left\{\frac{24 \pi G}{\gamma} \left(\Theta_\delta+\Theta_\sigma\right) t\right\},
\end{equation}
where $v_\theta(0)$ is the value of $v_\theta$ at a time $t=0$. Substituting this solution
in Eq.~\eqref{aeq}, we obtain the inhomogeneities:
\begin{equation}\label{asymptoticam}
  a_m = c_m e^{-i\omega_m},
\end{equation}
where $c_m$ is a complex integration constant and the phase $\omega_m$ depends
on time trough the variable $v_\theta$,
\begin{equation}
  \omega_m\equiv\frac{2\pi\gamma |m|}{\left(\Theta_\delta+\Theta_\sigma\right)}
  \left(\frac{3M}{2}\right)^{2/3} v_\theta^{2/3}.
\end{equation}
Introducing this solution into the definitions \eqref{eq:Hdefs}, we obtain the following
asymptotic behaviors,
\begin{subequations}\label{eq:H-as}\begin{align}
  H_0^\xi &= \sum_{m=-\infty}^\infty |m| |c_m|^2,  \tag{\ref{eq:H-as}} \\
  H_{\rm int}^\xi &= \sum_{m\neq 0}\frac{|c_m|}{|m|}
    \left\{|c_m|+|c_{-m}|\cos(2\omega_m-\phi_m-\phi_{-m} )\right\}, \notag
\end{align}\end{subequations}
where $\phi_m$ is the phase of the corresponding complex number $c_m$.

\section{Bianchi I}
\label{sec:B1}

As we have already explained,
Bianchi I is a particular case of the considered Gowdy spacetimes.
It can be described by the Hamiltonian constraint
\eqref{Hamiltonian} with vanishing inhomogeneities $a_m$ . Hence, the
effective dynamics for Bianchi I is given by the evolution equations
\eqref{eq:EOMs} with $H_0^\xi=H_{\rm int}^\xi=0$. In this case
$\Theta_\theta$ is also a constant of motion, which is determined by the
Hamiltonian constraint:
\begin{equation}
  \Theta_\theta = -\frac{\Theta_\sigma \Theta_\delta}{\Theta_\sigma + \Theta_\delta}.
\end{equation}

The system becomes thus sufficiently simple as to solve
the EOM analytically, namely,
\begin{subequations}\label{eq:B1-sols}\begin{align}
  v_i (t) &=  \frac{2}{3}\eta_i \,\Theta_i  \cosh[\alpha_i (t-t_0)+k_i],\\
  b_i(t) &= 2 \arctan \left[\tan\left(\frac{b_i(t_0)}{2}\right) e^{-\alpha_i (t-t_0)} \right],
\end{align}\end{subequations}
with the constants $k_i$ satisfying $\cosh{k_i}=1/|\sin{b_i(t_0)}|$, $\eta_i\equiv \sgn[\sin{b_i(t_0)}]$ and
\begin{equation}
  \alpha_i\equiv \frac{24\pi G}{\gamma} (\Theta_j+\Theta_k) , \quad (i\neq j\neq k).
\end{equation}
From these solutions, it is straightforward to see that bounces always
occur, since the hyperbolic cosine has a minimum. Making use of the conserved
quantities, it is easy to obtain the value of $v_i$ at the bounce point:
\begin{equation}\label{hombounce}
  v_i^{\rm bounce} = \frac{2}{3} \eta_i \,\Theta_i.
\end{equation}

\section{Analysis of the bounce}\label{sec:bounce}

In this section we focus our attention on the study of the bounce process. In particular,
we address the following questions:
\begin{enumerate}[(i)]
  \item Does the bounce persist in the presence of inhomogeneities?
  \item If that is the case, then is it generic? Does it happen always or are there
    trajectories reaching $v_i=0$, or at least entering the Thiemann-modified regime
    (where the quantum modifications to the inverse triad are important)?
  \item How does the presence of inhomogeneities affect the position of
    the bounce? If they happen to push the bounce point deeper towards the quantum
    regime (with respect to the analogous Bianchi I universe in vacuo), is there any lower bound
    on its position?
\end{enumerate}
Surprisingly, answering these questions does not require solving the EOM or carrying out
a detailed analysis of the dynamical trajectories. On the contrary, a lot of information can be
extracted just by studying the EOM, together with the Hamiltonian constraint, at the points
where the necessary condition for the bounce is satisfied
($\dot{v}_i=0$). These data are sufficient to define
the reflective surfaces on phase space which can never be crossed by any dynamical trajectory.
Whenever the dynamical trajectory hits such a surface, a bounce or recollapse necessarily occurs.
In this way, the question of whether the points $v_i=0$ can be reached boils down to the problem of elucidating
whether one can find a trajectory leading to any of those points while avoiding the commented reflective surfaces.

\subsection{Reflective surfaces}\label{sec:refl-surf}

\begin{figure*}[tbh!]
  \begin{center}
    $(a)$\hspace{3.2in}$(b)$
  \end{center}
    \vspace{-0.5cm}
  \psfrag{Logv}{$v_\theta/l_p^2$}
  \psfrag{H0}{$H_0^\xi/\hbar$}
  \includegraphics[width=7.3cm]{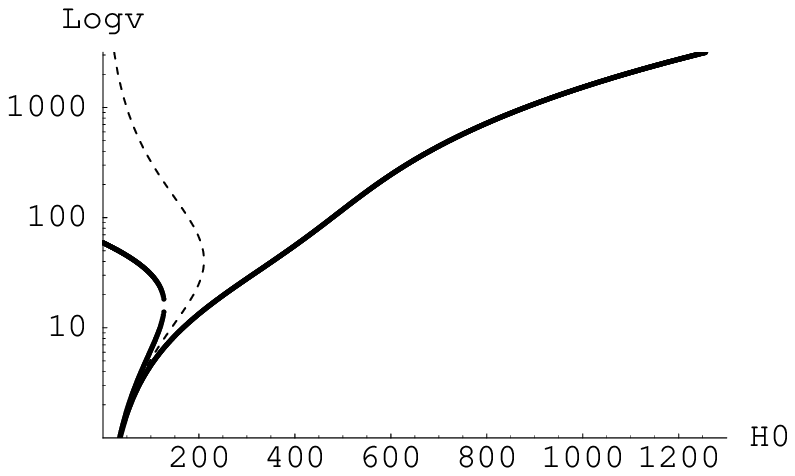}
  \psfrag{Logv}{$v_\theta/l_p^2$}
  \psfrag{H0}{$H_0^\xi/\hbar$}
  \includegraphics[width=7.3cm]{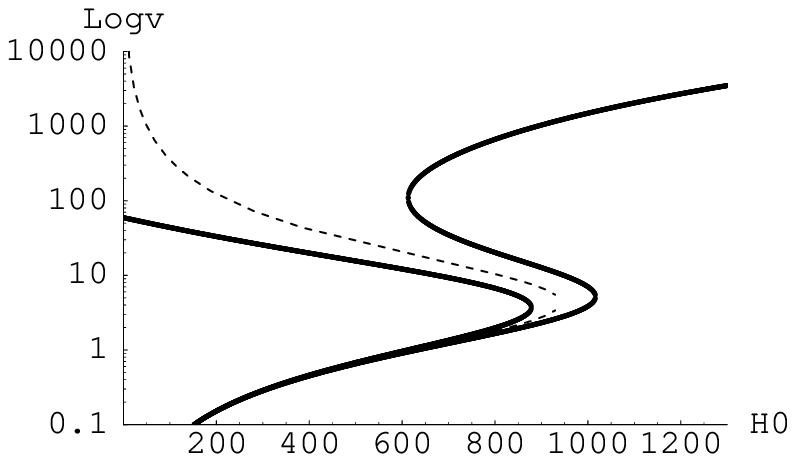}
  \caption{\label{roots} In this plot, we show all positive roots
    of Eq.~\eqref{sinbeq} in a logarithmic scale for the particular
    values $\Theta_\delta=800l_p^2$ and $\Theta_\sigma=100l_p^2$. In $(a)$
    $\alpha=0.01$, whereas in $(b)$ $\alpha=5\times 10^{-4}$.
    The dashed line corresponds to the section $F=\Theta_\delta\Theta_\sigma$.}
\end{figure*}

Let us start with the analysis of the bounce in the ``homogeneous'' directions
$\delta$ and $\sigma$. In this case, an inspection of the EOM provides the answer
almost straightforwardly. Indeed, the evolution equation for $v_\delta$ \eqref{vdeltaeq} implies
that, at the reflection point ($\dot{v_\delta}=0$), the function $\cos b_\delta$ vanishes. Applying this
to the constant of motion \eqref{eq:Theta-eff} and comparing with the homogeneous
result \eqref{hombounce} (for $i=\delta,\sigma$), we
immediately conclude that the value of $v_i$ at the bounce is exactly the same
as the value for its homogeneous counterpart (that is, the Bianchi I solution with
identical values of the constants of motion $\Theta_\delta$ and $\Theta_\sigma$).

This type of analysis is considerably more involved for the inhomogeneous direction $\theta$.
The value of $v_\theta$ at which the bounce occurs can be determined as
the largest (positive) root of the Hamiltonian constraint, treated as a function of $v_\theta$, upon
imposing the condition $\sin(b_\theta)\in\{\pm 1\}$. Indeed, setting this condition in
the constraint \eqref{Hamiltonian}
[with $\Theta_i$ expressed via Eq. \eqref{eq:Theta-eff}], and taking its square we get
\begin{equation}\label{sinbeq}
  \frac{9}{4}(\Theta_\delta +  \Theta_\sigma)^2 v_\theta^2=
  \left(F(v_\theta)-\Theta_\delta \Theta_\sigma\right)^2,
\end{equation}
where the function $F(v_\theta)$ is
\begin{equation}\label{defF}\begin{split}
  F(v_\theta) &\equiv 16 \pi^2\gamma^2 G \left(\frac{3 M}{2}\right)^{2/3} v_\theta^{2/3} H_0^\xi
  \\
  &+ \frac{G}{2}\left(\Theta_\delta + \Theta_\sigma\right)^2 \left(\frac{2}{3 M}\right)^{2/3}
    \frac{H_{\rm int}^\xi}{v_\theta^{2/3}}
\end{split}\end{equation}
and contains all the dependency on the inhomogeneities.

Given a dynamical trajectory, the variables $H_0^\xi$ and $H_{\rm int}^\xi$
are nontrivial functions of $v_\theta$. However one can assign to each trajectory the values
of $H_0^\xi$ and $H_{\rm int}^\xi$ at the bounce point. Then, on the family of all the dynamical
trajectories we can introduce classes of equivalence, where all the trajectories with the
same values of $H_0^\xi$ and $H_{\rm int}^\xi$ at the (first, if multiple are possible) bounce are identified as one class.
The parameter space of all these classes, coordinatized in particular by $H_0^\xi$ and $H_{\rm int}^\xi$,
is finite dimensional. On this space, the value of $v_\theta$ at the bounce
will be given by the largest root of Eq. \eqref{sinbeq}.
These solutions form precisely the reflective surfaces that we have mentioned above.

For the homogeneous case of Bianchi I, the function $F$ vanishes and we can
solve Eq. \eqref{sinbeq} analytically, obtaining the result that we already know [Eq. \eqref{hombounce}
for $i=\theta$]. In the most general situation, multiplying Eq.~\eqref{sinbeq}
by $v_\theta^{4/3}$ we get a fifth order polynomial $P$ for the variable
$v_\theta^{2/3}$, whose roots need to be found numerically.
Before embarking on this task, however, we note that we can obtain
interesting qualitative results analytically.

We first point out that there are two regions of clearly different
qualitative behavior, depending on
whether the function $F(v_{\theta})$ increases or decreases the value of the right-hand side
of Eq. \eqref{sinbeq}. Since $F$ is positive definite, this depends only on the sign
of the product $\Theta_\delta \Theta_\sigma$.

If it is negative, then the bounce (reflection) point is always higher than in the analogous homogeneous case.
In particular, this result ensures that the bounce occurs
before entering the region of small $v_\theta$, where our
approximation may not be valid.

On the other hand, if $\Theta_\delta \Theta_\sigma$ is positive,
the two terms on the right-hand side of Eq. \eqref{sinbeq} compete, and
the analysis is more complicated. In order to answer the questions posed at the beginning
of this section we have to analyze in detail the behavior of the roots of that equation.
As an illustration, in
Fig.~\ref{roots} we plot all the positive roots
for two particular cases (the negative roots are obtained from the positive ones just by a flip of sign).
As we will see below in more detail, the structure of the roots has
always this form.

Let us first coordinatize the space of equivalence classes of trajectories via the
following reparametrization of variables,
\begin{subequations}\label{eq:repar}\begin{align}
  |\Theta_\delta|\ &\equiv\ \frac{3v_h}{2\beta}  , \quad
  |\Theta_\sigma|\ \equiv\ \frac{3 v_h}{2(1-\beta)}  , \quad
  \beta\in[0,1]  , \notag
  \\ \label{defalpha}
  H_{\rm int}^\xi\ &\equiv\ \alpha H_0^\xi , \quad \alpha\in[0,2]  , \tag{\ref{eq:repar}}
\end{align}\end{subequations}
where $v_h=2|\Theta_{\theta}|/3$ is the position of the bounce
--in absolute value-- for the analogous Bianchi I universe
in the $\theta$ direction [see Eq. \eqref{hombounce}], and $\beta$ is a measure of
the asymmetry between the two homogeneous directions. The values $\beta=0,1$ correspond to the degenerated
cases when one of the homogeneous directions is suppressed or expanded
to infinity. The bounds on $\alpha$, on the other hand, are easily obtained from the definitions \eqref{H0def} and \eqref{Hintdef}
(taking into account that the mode number $m$ is always greater or equal than the unity).
The introduced choice of parameters simplifies the analysis considerably, because the only
noncompact coordinates are now $v_h$ and $H_0^\xi$.

To arrive to conclusive results, one needs to explore the entire space coordinatized as above, something that can be
done only numerically. Some analytical results can be however obtained in two asymptotic regions:
when $H_0^\xi$ is very small and when it is very large.

Suppose first that the inhomogeneities are small, that is, $H_0^\xi \ll 1$. Then, we can
neglect terms that are quadratic
in $H_0^\xi$ in the polynomial \eqref{sinbeq}. This leaves us with a fifth order polynomial in $v_\theta^{2/3}$
without free term. In this way, we obtain one root $v_\theta = 0$
and the remaining (reduced) polynomial is quartic, so that we can find its roots via analytic methods.
Apart from the zero, we obtain two complex and two real roots,
out of which, given the reality of $v_\theta^{2/3}$, only the latter two are relevant.
We expand them as a power series in $H_0^\xi$. As a result, we see that
one of them is zero up to order $(H_0^\xi)^2$ and the other, which is
the actual bounce point, is a little bit smaller than the position of the homogeneous bounce:
\begin{subequations}\label{eq:bounce-series}\begin{align}
  v_\theta^{\rm bounce} &= v_h - 16 \pi^2 G  \gamma^2\left(\frac{3 M}{2}\right)^{2/3}\frac{v_h^{5/3}}
    {\Theta _{\delta }\Theta _{\sigma }}H_0^\xi  \tag{\ref{eq:bounce-series}} \\
  &- 2G\left(\frac{2}{3M}\right)^{2/3}\frac{\Theta _{\delta } \Theta _{\sigma }}{9v_h^{5/3}}
    \alpha H_0^\xi+O\left[\left(H_0^\xi\right)^2\right]. \notag
\end{align}\end{subequations}
Note that this limit of small $H_0^\xi$ is equivalent to the
case of large $\Theta_\delta \Theta_\sigma$.

In the other region, i.e., for very large $H_0^\xi$, it is easy to
see that the only root of the equation tends to
\begin{equation}\label{largeH0}
  v_\theta^{\rm bounce}\approx \frac{2 M^2}{3}
  \left(\frac{16\pi^2\gamma^2 G}{\Theta_\delta+\Theta_\sigma}\right)^3
  \left(H_0^\xi\right)^3.
\end{equation}

Returning to the general situation,
the polynomial $P$ can be understood as
a second order polynomial in $H_0^\xi$. Finding its roots provides
the values of $H_0^\xi$ for which the bounce occurs at a given value of $v_\theta$.
These roots can be found analytically:
\begin{equation}
  H_0^\xi=\frac{3 v_\theta^{2/3}}{Y}(v_h\pm v_\theta),
\end{equation}
where we have defined
\begin{eqnarray}\nonumber
Y\!:=\!\left(\frac{2}{3M}\right)^{2/3} \!\!\alpha  G
 \left(\Theta _{\delta }+\Theta_{\sigma }\right)
+
\left(\frac{3M}{2}\right)^{2/3}
\frac{32  \pi ^2 \gamma ^2 G  v_\theta^{4/3}}{\Theta _{\delta }+\Theta_{\sigma }}.
\end{eqnarray}
It follows that
there are two positive roots for $v_\theta<v_h$ and one
positive root for $v_\theta>v_h$. Furthermore, for large $v_\theta$
the sole positive root is proportional to $v_\theta^{1/3}$, as we have
already shown via the asymptotic analysis \eqref{largeH0}.

Combining all these results, we get full information about
the number of real positive
roots of Eq. \eqref{sinbeq} that exist in different regimes.
As shown in Fig.~\ref{roots}, one branch
of roots starts at $v_h$ and decreases monotonously till it turns
around onto the branch of roots coming from zero. The sign
of $\partial v_\theta/\partial H_0^\xi$ can be determined analytically
in the following way. The derivative itself is given as the derivative
of the implicit function $P(v_\theta,H_0^\xi)=0$,
\begin{equation}
  \frac{\rd v_\theta}{\rd H_0^\xi}\ =\ -\frac{\partial P}{\partial H_0^\xi}\bigg/
  \frac{\partial P}{\partial v_\theta},
\end{equation}
and vanishes only when $\partial P/\partial H_0^\xi=0$. Substituting this
condition into $P(v_\theta)=0$ we
obtain $v_\theta=0$ as the only solution. As a consequence, in the region $v_\theta>0$ the
derivative $\rd v_\theta/\rd H_0^\xi$ cannot change sign
within each branch of roots. Therefore [and taking into account Eq. \eqref{eq:bounce-series}], the branch
starting from $v_h$ at $H_0^\xi=0$ is always strictly decreasing with
$H_ 0^\xi$.

By the very same method one can also prove that the line
$F(v_\theta)=\Theta_\delta\Theta_\sigma$ [with $F$ given in Eq.
\eqref{defF}] never intersects any of the roots of Eq. \eqref{sinbeq} outside
the point $H_0^\xi=0,v_\theta=0$.

\subsection{Minimum bounce point}\label{minimum}

In the previous subsection we gathered sufficient data
to characterize the reflective surfaces on phase space.
In particular, in the case $\Theta_\delta\Theta_\sigma<0$, we proved that the bounce point
is always larger than its homogeneous counterpart.
On the other hand, for $\Theta_\delta\Theta_\sigma>0$, and
identifying in the same equivalence class all the dynamical trajectories with identical
values of $H_0^\xi$ and $H_{\rm int}^\xi$ at the (candidate) bounce point, we showed that
the problem of finding a lower bound for the bounce can be recast as that of determining
the infimum of the largest real solution to Eq. \eqref{sinbeq}.
Now, have this largest solution been a continuous function of $H_0^\xi$ and $H_{\rm int}^\xi$,
its infimum would have provided exactly
the lower bound on the bounce position for the dynamical trajectories.
However, the function presents a discontinuity, a fact which complicates the situation.

\begin{figure}
  \psfrag{alpha}{$\alpha$}
  \psfrag{beta}{$\beta$}
  \psfrag{min}{$v_{\rm min}/v_h$}
  \includegraphics[width=7.3cm]{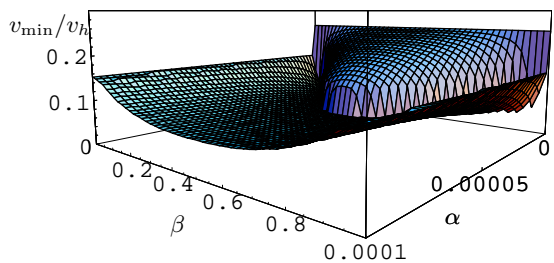}
  \caption{\label{fixvh} The ratio between the
    minimal root and the value of the fixed homogeneous bounce $v_h=10^5$ in terms
    of $\alpha$ and $\beta$.}
\end{figure}

\begin{figure}
  \psfrag{min}{$v_{\rm min}/v_{h}$}
  \psfrag{alpha}{$\alpha$}
  \psfrag{logvh}{$\log_{10} (v_h/l_p^2)$}
  \includegraphics{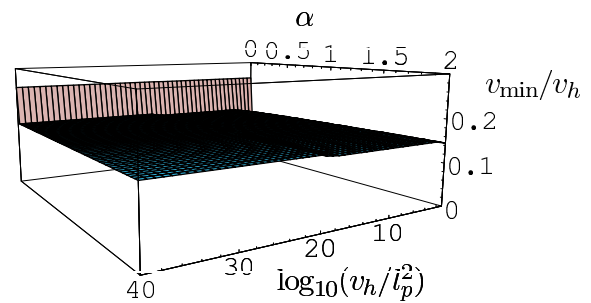}
  \psfrag{min}{$v_{\rm min}/v_{h}$}
  \psfrag{alpha}{$\alpha$}
  \psfrag{vh}{$(v_h/l_p^2)$}
  \includegraphics{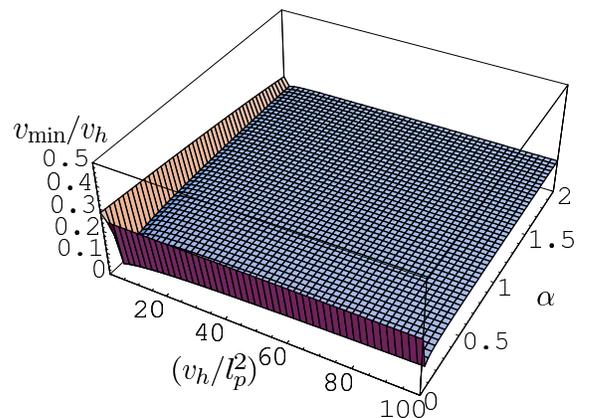}
  \psfrag{min}{$v_{\rm min}/v_{h}$}
  \psfrag{alpha}{$\alpha$}
  \psfrag{vh}{$(v_h/l_p^2)$}
  \includegraphics{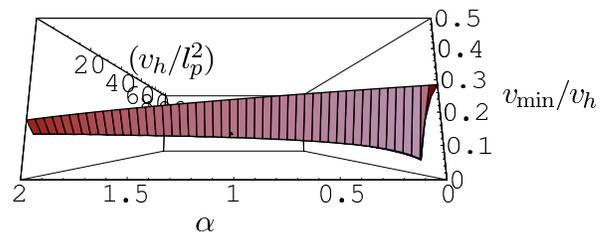}
  \caption{\label{fixbeta} The ratio $v_{\rm min}/v_{h}$ for a fixed
    value of $\beta=0.5$. In the first plot, $v_h$ appears in a logarithmic
    scale, but in a linear scale in the others.
    The second and third plots are in fact the same, only the perspective is modified.}
\end{figure}

\begin{figure}
  \psfrag{min}{$v_{\rm min}/v_{h}$}
  \psfrag{beta}{$\beta$}
  \psfrag{logvh}{$\log_{10} (v_h/l_p^2)$}
  \includegraphics{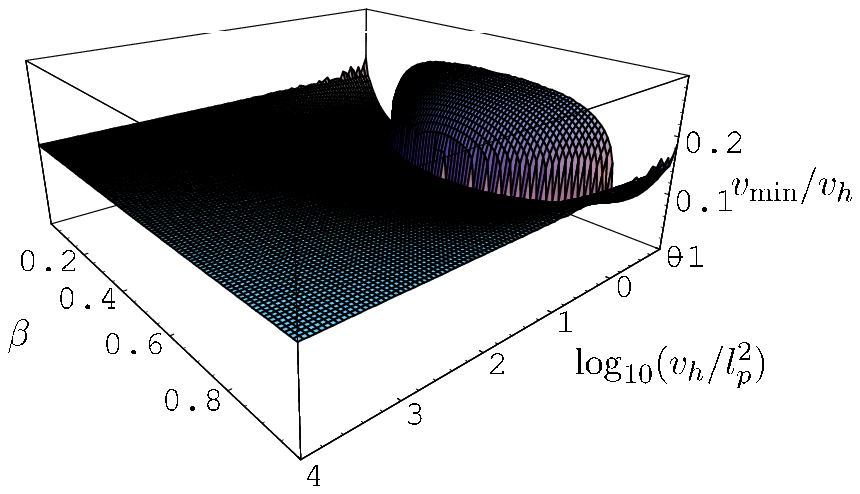}
  \psfrag{min}{$v_{\rm min}/v_{h}$}
  \psfrag{beta}{$\beta$}
  \psfrag{logvh}{$\log_{10} (v_h/l_p^2)$}
  \includegraphics{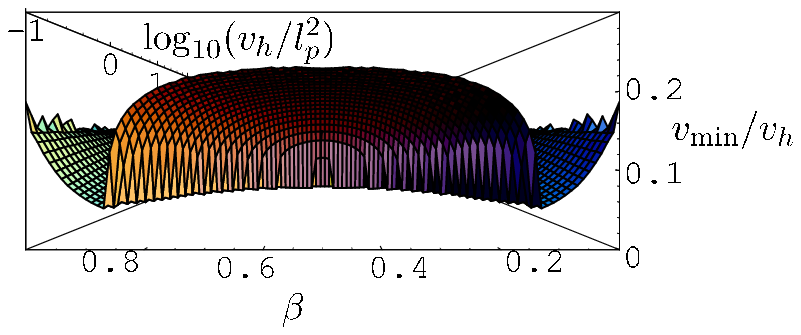}
  \caption{\label{fixalpha} The minimum bounce point $v_{\rm min}$ divided by its homogeneous
    counterpart $v_{h}$ for a fixed value of $\alpha=0.1$.}
\end{figure}

\begin{figure}
  \psfrag{min}{$v_{\rm min}/v_{h}$}
  \psfrag{vhom}{$v_h/l_p^2$}
  \includegraphics{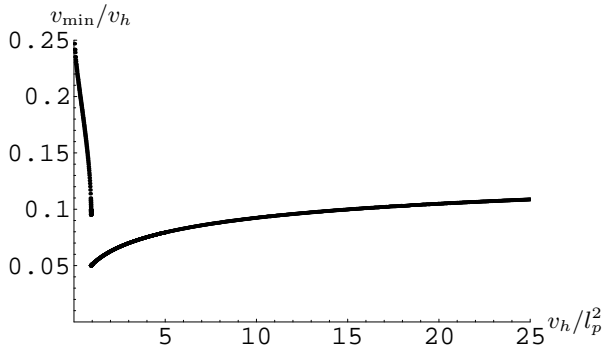}
  \caption{\label{fixalphabeta} The ratio $v_{\rm min}/v_{h}$ for fixed values of $\alpha=0.1$ and $\beta=0.5$.
    This is a section of the previous plots which shows the infimum of $v_{\rm min}\approx 0.05 v_{h}$.}
\end{figure}

Fortunately, an extensive numerical analysis
(described in detail in Sec.~\ref{sec:dyn})
shows that, on any given dynamical trajectory, the relative change
of $H_0^\xi$ is quite small. As a consequence, the trajectories
can be approximated by vertical lines on the plane $H_0^\xi$--$v_\theta$.
Moreover, the roots of Eq. \eqref{sinbeq} change slowly with $H_{\rm int}^\xi$.
In this situation,
a lower bound on the position of the (first) bounce (if multiple bounces are possible)
for fixed $v_h$ and $\beta$
is indeed given by the corresponding infimum of the largest solution to Eq. \eqref{sinbeq}.
Owing to our approximations, nonetheless, the equality between the former bound and this infimum
is also approximate only. However, it is not necessary to determine the bound exactly.
Its order of magnitude is enough to establish the kind of qualitative results we are interested
in (furthermore, our result will be checked in Sec.~\ref{sec:montecarlo} by performing a numerical analysis of large
populations of dynamical trajectories).

As we have seen, for small $H_0^\xi$ the
bounce point is smaller than its homogeneous counterpart,
and this value decreases with $H_0^\xi$. This provides the largest root of Eq. \eqref{sinbeq}
until the number of positive roots changes. Beyond this discontinuity, the new largest root belongs to the branch
that starts at zero for vanishing $H_0^\xi$ (as can be checked in Fig. \ref{roots}), and becomes
an increasing function of the inhomogeneities $H_0^\xi$. Therefore, the reflecting boundary
exists for all values of $H_0^\xi$, and the infimum of the desired function occurs exactly
at the commented discontinuity. We accordingly adapt our search methods to these considerations.

On a general level, the discussion reduces to study the function $v_B/v_h$, where $v_B$
is the largest root of Eq. \eqref{sinbeq} for fixed parameters
$(H_0^\xi,\alpha,\beta,v_h)$. The problem is split into several steps as follows.
First we fix the values of all the parameters except $H_0^\xi$ and
find the minimum (or rather infimum, because it happens at a discontinuity)
of $v_B/v_h$ with respect to $H_0^\xi$, which we call $v_{\rm min}$.
Next we analyze the properties of $v_{\rm min}(\alpha,\beta,v_h)$.
Since this analysis must be performed numerically, we choose
natural units with $G=1$ and $l_p=1$, whereas for the Immirzi
parameter we take the value deduced from the computation of
black hole entropy \cite{dl-gamma,*m-gamma}, i.e., $\gamma=0.23753295\ldots$

In order to find the value of $H_0^\xi$ at which
$v_{\rm min}$ is attained,
we first note that it must be
smaller than $H_0^*$, which is defined as the maximum value of $H_0^\xi$ allowed by the implicit function
\begin{equation}
  F(v_\theta,H_0^\xi) = \Theta_\delta\Theta_\sigma
\end{equation}
for positive $v_\theta$.
This can be clearly seen in Fig.~\ref{roots}
(see also the discussion at the end of Sec.~\ref{sec:refl-surf}
for more precise conclusions). Therefore, to
find $v_{\rm min}$ it is enough to explore numerically the
interval $H_0^\xi\in[0,H_0^*]$.

This numerical analysis was performed in the following way. The domain
$H_0^\xi\in[0,H_0^*]$ was divided in
$1000$ uniform intervals and the largest root of Eq.~\eqref{sinbeq} was
found at each point $n H_0^\xi/1000$, for $n=0,\dots, 1000$.
The minimum of those roots was chosen, with its corresponding value
of $H_0^\xi\equiv H_0^{\rm min}$, and the study was then
repeated in the interval $[H_0^{\rm min} -H_0^*/1000, H_0^{\rm min} +H_0^*/1000]$.
In Figs.~\ref{fixvh}, \ref{fixalpha}, \ref{fixbeta}, and \ref{fixalphabeta}
we show the behavior of the function $v_{\rm min}(\alpha,\beta, v_h)/v_h$ with respect
to the different parameters. In each plot, one of the three parameters
$(\alpha,\beta, v_h)$ is fixed.

The main result found is that the value of the minimum bounce point
is bounded from below by $v_{\rm min}\approx 0.05 v_h$, as it is evident from
Fig.~\ref{fixalphabeta}. This
bound is approached at values of $\alpha$ and $v_h$ that are small. In fact
for large $v_h$ and/or $\alpha$, the ratio $v_{\rm min}/v_h$
tends to a number close to $0.15$. In conclusion, we have shown for the inhomogeneous case that
$v_\theta$ at the bounce is always larger than a 5\%
of its homogeneous counterpart, or in other words, it is never two or more
orders of magnitude smaller.

Let us emphasize that the analysis performed in this
section does not actually solve the EOM.
Instead, we only provided reflective boundaries on phase
space. Nonetheless, since these boundaries are discontinuous,
we can not exclude that a dynamical trajectory may slip between different
branches of roots and reach the singularity $v_\theta=0$ in that way.
However, this possibility would require an extreme fine tuning.
Furthermore, it is worth noting that such a possibility would
require that the value of $H_0^\xi$ on the trajectory
decay to zero in the evolution. This implies that, in any case, the system would need
to homogenize itself in order to reach $v_{\theta}=0$.
This behavior might then be interpreted as a particular realization of the BKL conjecture \cite{bkl},
restricted to the case of the Gowdy spacetime investigated in this article.

Moreover, the above situation is possible only on trajectories for which all
constants of motion $K_m$ [defined in Eq. \eqref{km}] vanish.
An example of a near-critical dynamical trajectory of the system can be found in
Fig.~\ref{critical}, where the
initial data are chosen so that the bounce happens near the discontinuity in the largest root of Eq. \eqref{sinbeq}.
In this case, the system bounces and recollapses back and forth several times
before escaping to the region of asymptotically large $v_\theta$.

\section{Numerical study of the dynamics: the bounce}\label{sec:dyn}

Despite the fact that a lot of information about the system can be extracted just from
the Hamiltonian constraint and the EOM without integrating them,
a full answer to most of the questions posed
at the beginning of this paper requires the numerical determination of the dynamical trajectories.
In this section we present the numerical techniques that we have used to solve the EOM,
and display and discuss explicit examples of solutions.
Here, we focus our attention on the behavior of the homogeneous variables (namely $v_i$). The analysis of the
behavior of the inhomogeneity modes is postponed to the next section, since it requires more specific
methods.

Let us start with the initial value problem formulation and the description of the integration method.

\subsection{The initial value formulation}\label{formulation}

The analysis of the asymptotic behavior of the solutions of Sec.~\ref{sec:asympt} revealed that,
in the large volume regime, each of the inhomogeneities $a_m$ becomes an oscillating function of constant amplitude,
with a time-dependent frequency that is proportional to $v_\theta^{2/3}$ \eqref{asymptoticam}.
Since one enters that regime when the solutions to Eq. \eqref{eq:EOMs} are constructed, these oscillations
require elaborated integration methods, forcing one to
refine the integration step and increasing the computational time significantly.
To avoid this difficulty, we implement two changes in the formulation of the system.
First, we change the time $t$ to a new time $t'$ defined as,
\begin{equation}\label{newt}
  \frac{\rd}{\rd t'}\equiv \frac{1}{|v_\theta|^{2/3}}\frac{\rd}{\rd t}.
\end{equation}
In practical terms, this is easily done by dividing
the right-hand side of Eqs. \eqref{eq:EOMs} by $|v_\theta|^{2/3}$.
With the new choice of time the asymptotic behavior of the inhomogeneity
modes takes the following form:
\begin{equation}\label{fromatoc}
  a_m = c_m e^{-i\tilde\omega_m t'},
\end{equation}
where the frequency $\tilde{\omega}_m$ is
\begin{equation}
  \tilde\omega_m\equiv 32\pi^2 G |m| \left(\frac{3 M}{2}\right)^{2/3}.
\end{equation}
This frequency depends on the absolute value of $m$ but not on time.
As a result, now each inhomogeneity mode oscillates with a constant frequency.

Next, it is convenient to introduce
a change of variables from $a_m(t')$ to $c_m(t'):=a_m(t')e^{i\tilde\omega_m t'}$.
The asymptotic relation \eqref{fromatoc} implies then that the new variables $c_m(t')$
approach asymptotically the constants $c_m$. This allows one to significantly improve
the efficiency of the numerical integration, because the oscillatory behavior is removed in
the regime of large $v$, permitting one to adopt larger integration steps there.

As we explained in Sec. II, the system of equations that we want to solve numerically
has several conserved quantities: the Hamiltonian and momentum constraints
[\eqref{Hamiltonian} and \eqref{constraint} respectively], the constants $K_m$ [given in Eq. \eqref{km}],
$\Theta_\delta$, and $\Theta_\sigma$. In principle, one can solve the constraints they impose
on the system, decreasing in this way the number of variables. Here, however,
we use a free-evolution scheme and employ the
conservation of these constants of motion as a (quantitative) test on the
correctness of our numerical method. Therefore, the equations that
we want to integrate numerically are those for the variables
$\{v_i, b_i\}$ [(\ref{vthetaeq}-\ref{bthetaeq}) and
(\ref{vdeltaeq}-\ref{bdeltaeq})] in terms of the new time
variable $t'$. In addition, the equations \eqref{aeq} are replaced with the evolution
equations of the new inhomogeneous variables $c_m$, which are
\begin{equation}\label{eq:cm-evo}\begin{split}
  \frac{\rd c_m}{\rd t'} &= - \frac{i G}{\gamma^2 |m||v_\theta|^{4/3}}
    \left(\frac{2}{3M}\right)^{2/3}  \\
  &\hphantom{=}
    \times \left(\Theta_\delta + \Theta_\sigma\right)^2
    \left( c_m^{} + c_{-m}^* e^{2 i \tilde\omega t'} \right).
\end{split}\end{equation}

To compare the solutions of these effective equations with their counterpart in General Relativity, we also need to
solve the classical EOM of our Gowdy spacetimes. In practice, these can be obtained from the effective equations
introduced above
just by making the transformations $\{\sin b_i\rightarrow b_i, \cos b_i\rightarrow 1\}$
(see the discussion in Sec.~\ref{sec:eoms} for the validity of this method).

As initial conditions for our simulations, we select data corresponding
to classical universes. In particular, we always start the simulation at large values of
the variables $v_i$ and small values of the variables $b_i$.
The crucial factor responsible for the variety of studied solutions is found in
the initial conditions for the inhomogeneity modes.
On the one hand, different cases come from a different number of excited modes.
On the other hand, the amplitude of the modes can also vary
from case to case.

The constraints \eqref{Hamiltonian} and \eqref{constraint}
must be satisfied, in particular, on the initial slice; hence they must be taken into account
in the construction of admissible sets of initial data.
The exact construction is as follows. First, we build data with just
$n$ excited modes, that is,
the only nonvanishing modes $c_m$ are those with $|m|\leq n$,
for some fixed number $n$.
Except for
${\rm Im}(c_n)$, the values of all the initial data $c_m$ are constructed via
a random number generator, with a probability density
function given by a Gaussian of vanishing mean value.
Finally, we obtain the remaining piece of data, ${\rm Im}(c_n)$, by solving the momentum constraint
\eqref{constraint}.

The amplitudes of these modes are of the order of the standard deviation
of the probability distribution function, which is the remaining freedom in the construction.
We exploit this freedom, changing this deviation in several ways
(see Appendix~\ref{app:plots} for details) in order to check the robustness of the result.
With the same aim, in some simulations we actually use the flat distribution instead of the Gaussian one,
but this choice does not lead to any relevant difference in the results.

Given the initial data for the inhomogeneities obtained in this manner, we next
fix the remaining free parameters:
$(v_\delta, b_\delta)$, $(v_\sigma, b_\sigma)$, and $v_\theta$.
The initial value of the remaining variable $b_\theta$ is then determined by the
Hamiltonian constraint \eqref{Hamiltonian}.

It is important to recall that we are interested only in initial data which
provide universes in the classical evolution regime. Therefore,
after constructing the data, we have to verify that $v_ib_i \simeq v_i\sin b_i$ for all
directions $i$, up to the desired accuracy
(this is a condition additional to the ones discussed above).

Since the universe bounces in the effective trajectory, in order to
compare it with its classical counterpart we have to solve the classical
EOM for two cases. Namely, the case with the same initial data
and the case where those data are provided by the
effective trajectory at large distant times in the future.
This latter case gives a good approximation to
the classical universe that the effective trajectory
approaches after the bounce.
These two classical universes are related by a flip of
sign in the quantities $\Theta_\delta$ and $\Theta_\sigma$.

Once the initial data are selected, the system of EOM is integrated via the
built-in \emph{Mathematica} adaptive function \texttt{NDSolve},
which by default uses an Adams method and a Gear backward difference method, switching
dynamically between them \cite{math-man}. For our
simulations, specifically, we set the error control options to \texttt{AccuracyGoal}$\rightarrow 10$,
\texttt{PrecisionGoal}$\rightarrow 12$, and \texttt{WorkingPrecision}$\rightarrow 25$.
Since only a finite number of
modes has been excited initially and the linear equations \eqref{eq:cm-evo} are homogeneous within
each set of modes with identical values of $|m|$,
only a finite number of EOM needs to be integrated.

Let us end this subsection with a remark. In order to overcome technical difficulties
(infinite number of EOM) only a finite number of modes has been allowed to be excited.
One could then argue that the system has been effectively truncated to one with only a finite
number of degrees of freedom, and thus its behavior might be significantly
different from its field-like counterpart. Even so, and at least
from our effective perspective, both systems (the truncated one and the
infinite dimensional system) should behave similarly. This is supported
by the following argument. Since the value of $H_0^\xi$ provided on the initial section
is finite, the series \eqref{H0def} which defines it must converge at least near the initial time.
This implies that there exists a finite positive integer $l$ such that $|a_m|>|a_{m+1}|$ for all $|m|\geq l$.
On the other hand, the form of the EOM ensures that the growth of each individual mode
cannot be faster than exponential. Hence the quantities $H_0^\xi$ and $H_{\rm int}^\xi$ may diverge only
if the amount of the relative change of $|a_m|$ across the bounce increases sufficiently fast with $|m|$.
However, as we will show in Sec.~\ref{sec:montecarlo}, statistically there is no detectable evidence that this
change depends on the value of $|m|$ at all.

Furthermore, the inhomogeneities interact only within pairs $(m,-m)$, and their amplitude
can be ``pumped up'' only through the backreaction of the homogeneous background. As we will see in
Sec.~\ref{sec:inh}, this process occurs only near the bounce and does not depend on the number of excited
modes (as the EOM for the homogeneous variables involve only the total energies $H_0^\xi$ and $H_{\rm int}^\xi$).
As a consequence, one can extrapolate the commented results about the dependence of the amplification on $|m|$
to the case of an infinite number of modes.
This indicates that a change in the mode energy distribution that is sufficiently strong as to destroy the convergence
of $H_0^\xi$ occurs with vanishing probability.
Therefore, the contribution to the system dynamics of all the modes with $|m|$ larger than a certain threshold
ought to be negligible.

\subsection{The dynamical bounce}\label{sec:dyn-bounce}

\begin{figure*}
  \psfrag{s}{$\log_{10}(v_\sigma/l_p^2)$}
  \psfrag{t}{$\log_{10}(v_\theta/l_p^2)$}
  \includegraphics[width=7.3cm]{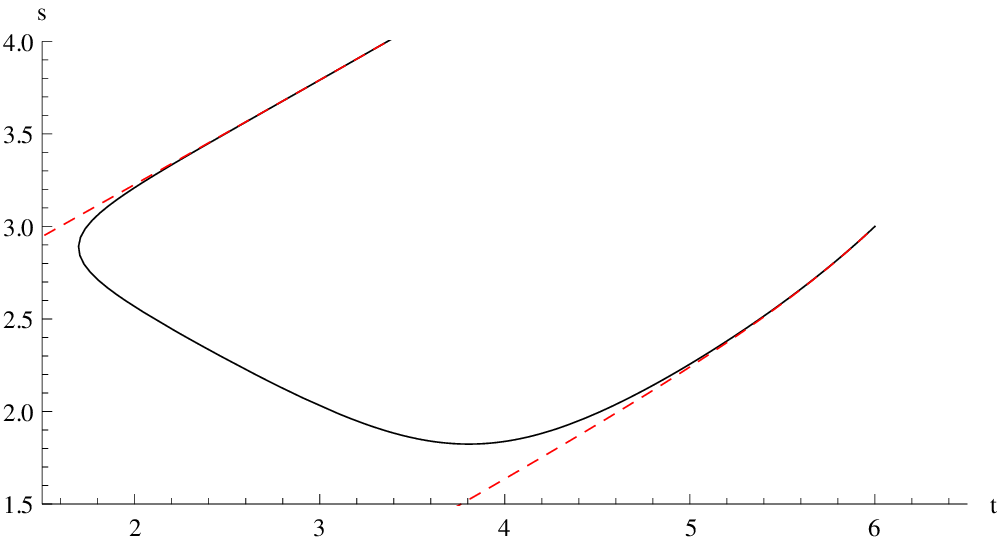}
  \psfrag{t}{$\log_{10}(v_\theta/l_p^2)$}
  \psfrag{d}{$\log_{10}(v_\delta/l_p^2)$}
  \includegraphics[width=7.3cm]{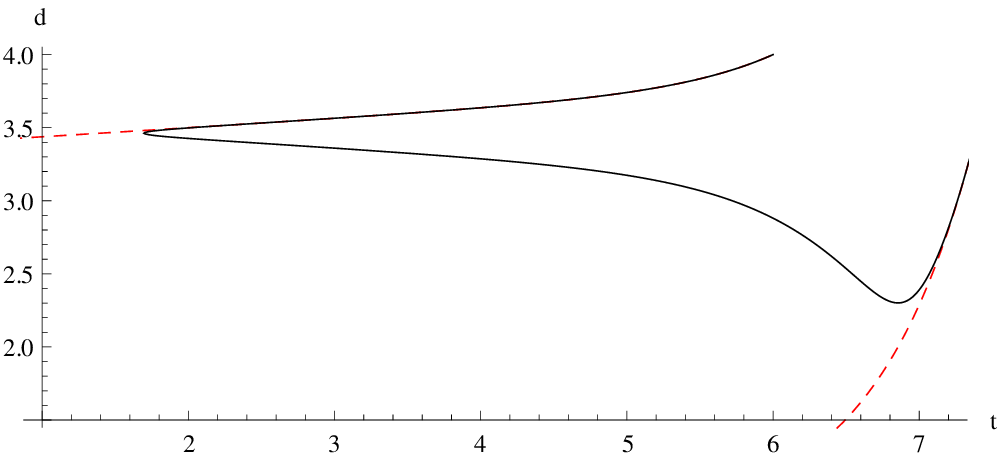}
  \caption{\label{logvlogv5} The black (continuous) curves are dynamical effective trajectories
    which, after bouncing in all the three directions, converge to
    the classical trajectories displayed in (dashed) red.
    The initial data are
    $v_\theta=10^6l_p^2$, $v_\delta=10^4l_p^2$, and $v_\sigma=10^4 l_p^2$, with constants of motion
    $\Theta_\delta=300l_p^2$ and $\Theta_\sigma=100l_p^2$. Only five inhomogeneity modes have been excited.
    In order to obtain initial conditions
    for those inhomogeneities, the width of the Gaussian random number generator
    has been chosen to be $10^{-1}$. This has led to the initial
    values $H_0^\xi\approx 0.3846\hbar$ and $H_{\rm int}\approx 0.02719\hbar$ ($\alpha\approx0.0707$).
    The two classical trajectories are related by a flip of sign in the constants $\Theta_\delta$
    and $\Theta_\sigma$.
  }
\end{figure*}

As a way to present the results of the numerical analysis of the dynamical
trajectories, let us give an example of a typical solution and discuss its behavior.

In Fig.~\ref{logvlogv5} we show a generic example of regular
evolution in a logarithmic plot on the $(v_\theta,v_\delta)$ and
$(v_\theta,v_\sigma)$ planes of phase space.
This specific plot represents a system
with five modes. The Gaussian width $\sigma$ of the random number generator
that provides us with the initial conditions for the inhomogeneities
has been chosen equal to $10^{-1}$, and leads to the approximated
values $H_0^\xi\approx 0.3846\hbar$ and $H_{\rm int}\approx 0.02719\hbar$
(this makes $\alpha\approx 0.0707$). We observe that the effective trajectory,
represented by a black (continuous) line, asymptotically connects two different classical
trajectories. Namely, the effective trajectory converges to two distinct classical ones in its distant
past and future, respectively. As we have already mentioned, these two classical
trajectories are related by a flip of sign in $\Theta_\delta$ and $\Theta_\sigma$ which, for this particular example, take
the values $\Theta_\delta=300 l_p^2$ and
$\Theta_\sigma=100 l_p^2$.
The evolution begins at $v_\theta=10^6 l_p^2$,
$v_\delta=10^4 l_p^2$, and $v_\sigma=10^4 l_p^2$. In the $(v_\theta,v_\delta)$ plane,
the effective trajectory follows a classical one corresponding to a contracting universe
almost until a bounce in $v_\theta$ happens. Near this bounce, the solution departs
from the classical (diverging) trajectory and, very quickly after
$v_\delta$ bounces, approaches the new expanding classical trajectory.

The numerical simulations have been performed for various values
of the inhomogeneities, with data
generated by a Gaussian random number generator of changing width.
This leads to a population of different trajectories in phase space,
for which the behavior of the bounce agrees with the results of
the analysis of Sec.~\ref{sec:bounce}. With exception of the near-critical trajectories,
the behavior is qualitatively the same as described for the example
presented in Fig.~\ref{logvlogv5}.

\begin{figure}
  \psfrag{vtheta}{$v_\theta/l_p^2$}
  \psfrag{H0}{$H_0^\xi/\hbar$}
  \includegraphics[width=7.3cm]{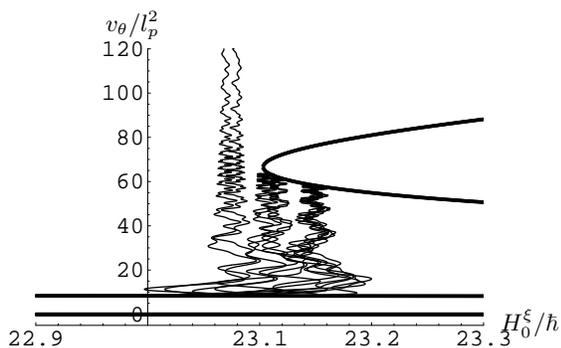}
  \caption{\label{critical} An example of the near-critical behavior
    of the evolution near the throat formed by different branches of roots.
    The system bounces and recollapses
    several times before going back to the region with
    large values of $v_\theta$.}
\end{figure}

To complete the analysis of this section, let us consider now
a near-critical solution. An example of such a kind of solution
is shown in Fig.~\ref{critical}. In that particular simulation,
we have chosen initial data with $K_m=0$, a necessary condition for the system to be
potentially able to reach the singularity.
The plot shows a zoom of the dynamical trajectory, presenting the variation of $H_0^\xi$
in terms of $v_\theta$ near the mentioned point. During the evolution,
the system becomes trapped between the two ``leaves" of the reflective surface.
It then repeatedly bounces and recollapses, upon hitting the lower
and upper leaves corresponding to the different branches of roots
of Eq. \eqref{sinbeq}.
After several of these cycles, the system is finally able
to escape from this region to the sector of classical regime.
It is extremely difficult to say something general and precise about the
evolution of the system near this ``throat", formed by the distinct branches of roots.
Actually, this evolution
is highly sensitive even to very small changes of the mean value
of the inhomogeneities during the bounce.
In principle, the possibility exists that
the trajectory may even go all the way ``down the throat'' and reach
the singularity in an infinite sequence of bounces
and recollapses. However, such critical trajectories require a
fine tuning of parameters; thus (apart from being unphysical
due to the own limitations of the effective treatment) they should
be disregarded based on statistical considerations.

\section{Numerical study of the dynamics: the inhomogeneities}\label{sec:inh}

Once we have established that the bounce persists, it is natural
to ask how the structure of the inhomogeneities changes across it. This section explores this
issue. In particular, we will focus our discussion on the amplitudes $|a_m|$ of the different
inhomogeneity modes, since these
quantities contain physical information about the energy of those gravitational waves modes.

\subsection{General description}\label{sec:num-gen}

\begin{figure}
  \begin{center}
     $(a)$ \hspace{2cm} $(b)$
  \end{center}
  \vspace{-0.6cm}
  \psfrag{vth}{$v_\theta/l_p^2$}
  \psfrag{|a|}{$|a_m|/\sqrt{\hbar}$}
  \includegraphics[width=7.3cm]{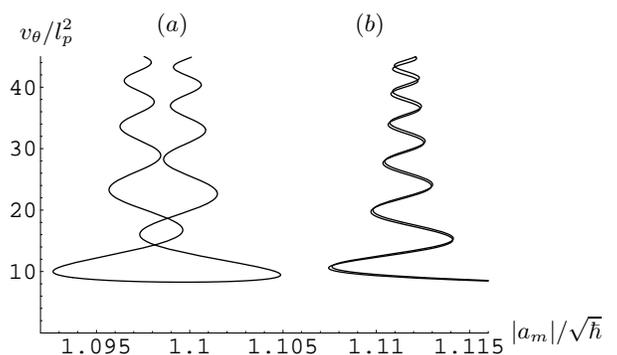}
  \caption{\label{amvth}
    Typical evolution of the amplitude $|a_m|$ of two modes of the inhomogeneities. During the
    evolution, away from the bounce, its mean value (obtained by subtracting the oscillations) is conserved.
    However, we see two distinct types of
    behavior for the change of this mean value
    trough the bounce in $v_\theta$. In case $(a)$, the mean value
    changes by an amount which is almost one amplitude of the oscillations,
    whereas it remains
    practically unchanged in case $(b)$.
  }
\end{figure}

Let us first study the behavior of the amplitude of a single inhomogeneity mode.
A typical dynamical trajectory is displayed in
Fig.~\ref{amvth}, the evolution being represented with respect to $v_\theta$.
As one can see, before and after the bounce (in $v_\theta$) the trajectory of $|a_m|$
describes sinusoidal oscillations about a constant (conserved) mean value.
The amplitude of these oscillations behaves like $v_{\theta}^{-\epsilon}$ for certain
positive number $\epsilon$. In particular, for large values of $v_\theta$, $|a_m|$ tends to a
constant value $|c_m|$, as proves
the asymptotic solution \eqref{asymptoticam}.
On the other hand, the amplitude of the oscillations is amplified as
the trajectory approaches the bounce, reaching the maximum there.
At that point, the oscillatory behavior is interrupted temporarily.
This may result in a change of the mean value of the considered mode.
In particular, Fig.~\ref{amvth} shows
the two different possible behaviors of the modes
when crossing the bounce. In this example, the mean value increases in one case
by a full amplitude of the oscillations, whereas in the other case the mode traces
back its pre-bounce trajectory almost perfectly after the bounce,
and thus its mean value is approximately preserved.

In order to understand the physical consequences of these different
behaviors, we have integrated the evolution for
a set of initial data where only the phase
of a given mode is allowed to vary, namely, by a phase shift $\varphi$.
We have then measured the change of the mean value of this mode after the bounce,
treated as a function of the phase shift. Technically, the mean value
has been identified with the value at the point where the second derivative
of the mode amplitude $|a_m|$ vanishes.

\begin{figure*}
  \psfrag{dam}{$\Delta |a_m|$}
  \psfrag{phi}{$\varphi$}
  \includegraphics[width=7.3cm]{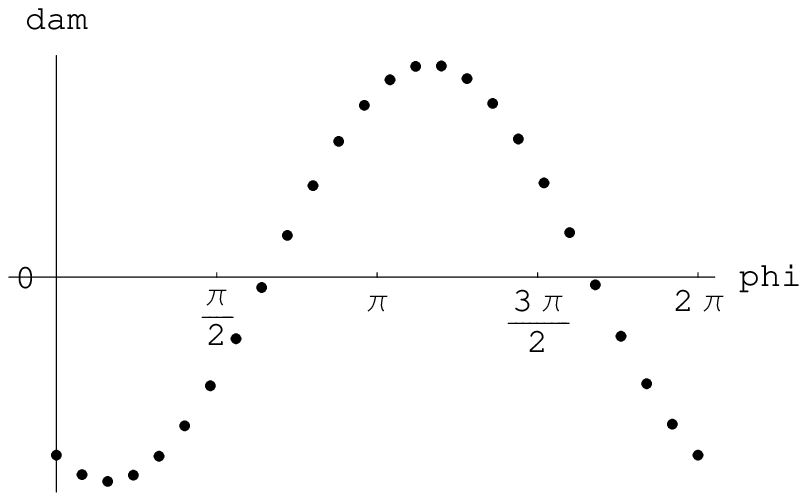}
  \includegraphics[width=7.3cm]{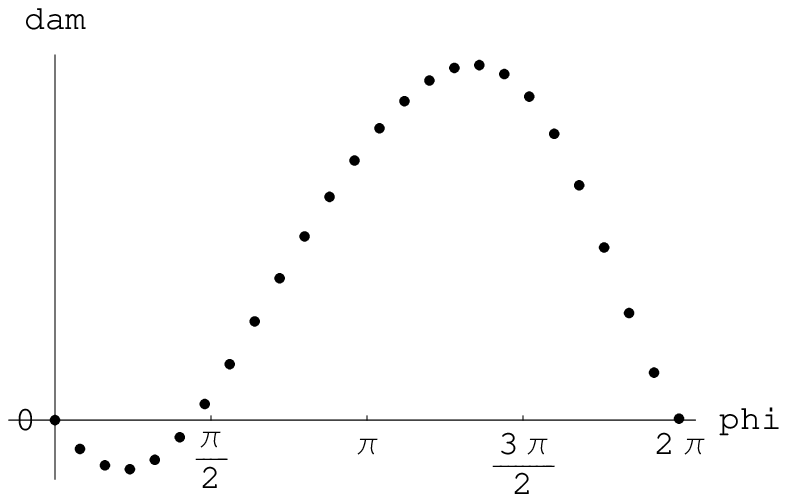}
  \caption{\label{dephase} The dependence of the change
    of the amplitude through the bounce on the initial
    phase shift of the mode.}
\end{figure*}

This study again reveals two distinct types of behavior, presented
in Fig.~\ref{dephase}. In the first case, that corresponds to the situation where
the oscillations of $|a_m|$ near the bounce are small in comparison to its mean value,
the change in this mean value is antisymmetric under a rotation of the
initial phase shift, $\varphi\rightarrow\varphi+\pi$.
Hence, for large samples, the mean value of the amplitude of the inhomogeneities,
and thus of $H_0^\xi$, is conserved through
the bounce. The second case corresponds to the opposite situation.
The sinusoidal function is antisymmetric around certain
positive axis; thus it has a positive integral. This implies that, in this case, the inhomogeneities
are statistically amplified.
We will further investigate this effect in the next section,
carrying out a systematic statistical analysis.

\subsection{Statistical analysis of the energy distribution}\label{sec:montecarlo}

In modern cosmology, a great deal of attention is paid to the analysis of
the energy distribution of the inhomogeneities, with the phases of the modes
being often ignored because, essentially, they do not affect the observations.
Furthermore, in the class of cosmological
models considered here, those phases influence the universe dynamics only
through the quantity $H_{\rm int}^{\xi}$, which
is suppressed with respect to $H_0^\xi$ (which does not depend on the phases)
by a factor $v_\theta^{2/3}$.
As a consequence, in the far future and past, where the universe is in a
classical regime, their effect is negligible.

One has to recall that they are still part of the initial data (see Sec.~\ref{formulation}),
so they are necessary to determine the dynamical trajectory in a unique way.
However, it is natural to ask
what can be said about the evolution once they are ignored.
In particular, given data  in the distant past determining the
energies of the gravitational wave modes (encoded just in $|a_m|$), one can ask about
their value and properties in the distant future. Since we are neglecting part of the initial data,
the nature of the analysis and of the results necessarily becomes statistical.

With this motivation in mind, in this section we are going to analyze
the statistically averaged change of energy of the individual modes
during the transition from the distant past to the distant future.
This is a basic piece of information, necessary to assess
the changes in the energy distributions.

To achieve this goal, we employ a Monte-Carlo analysis, that is, we
carry out a large number of numerical integrations of the system
using different random initial data in order to cover a portion
of phase space as large as possible.
Since the statistical behavior has proven not
to depend on the number of excited modes (as one can check by inspection),
we have performed the bulk of the simulations just for the case in which only
the first five modes are excited, i.e., we assume $a_m=0$ for all $|m|>5$.
The results of these simulations, nonetheless, have been verified on much smaller samples
of data but with already twenty modes excited.

In addition, to ensure the robustness of the results, the simulations have been performed
for six different forms of the energy distributions. For each individual trajectory,
the construction of the set of initial data has been specified along the lines explained in
Sec.~\ref{formulation}. In more detail:

$\bullet$ The value of the constants of motion $\Theta_\delta$ and $\Theta_\sigma$
have been given in terms of the partly compactified coordinates $v_h$ and $\beta$, as defined
in Eq. \eqref{defalpha}, and whose values have been chosen by a random number generator with flat
distribution. In particular, the value of the anisotropy factor $\beta$ has been selected from the interval
$[0.1, 0.9]$ to exclude the uninteresting near-degenerate cases from the analysis.
The value of $v_h$ has been chosen as the exponential of a number (again provided by a random
number generator of flat distribution) from certain interval. This interval depends on the data set
(see the specification of the inhomogeneities below) and is equal to: $[0, \log 10^9]$ for
cases $(i)$-$(iv)$, and $[0, \log 10^5]$ for $(v)$-$(vi)$.

$\bullet$
The initial values of the inhomogeneities [all but ${\rm Im}(a_5)$,
which is determined by the diffeomorphism constraint]
have been obtained (real and imaginary part separately) using
a random number generator with a Gaussian distribution of vanishing mean value.
The width $\sigma$ of this Gaussian has been selected with the use of six different algorithms, which in fact define
our data sets. For each of these algorithms, $\sigma$ has been specified as follows:
$(i)$ $\sigma$ is a random number (again with flat distribution) from the interval $[0,1000]$,
$(ii)$ $\sigma=100$,
$(iii)$ $\sigma=0.1 v_h^{2/3}$,
$(iv)$ $\sigma=0.5 v_h^{2/3}$,
$(v)$ $\sigma=2 v_h^{2/3}$, and
$(vi)$ $\sigma=v_h$.

$\bullet$ Given the above data, we have solved Eq.~\eqref{sinbeq} to estimate
the bounce point, choosing the initial value of $v_\theta$ as $10^7$ times the
obtained value. In this way, it is assured that the value of $v_\theta$ is
large enough as to guarantee that the initial point lays in the classical regime.

In the last four of the above sets of initial data for the inhomogeneities,
we have related the value of $\sigma$
with the homogeneous bounce point $v_h$. The reason for this is that,
in a universe with large $v_h$, the inhomogeneities should
be larger in order to have a similar effect in its dynamics.
This is motivated by the fact that
the amplitude of the inhomogeneities (or equivalently of $H_0^\xi$) by itself
is not a good measure of the departure of a universe from homogeneity.
For that purpose one should rather consider the relative values of the terms
that contain the homogeneous
and the inhomogeneous contributions in
the Hamiltonian constraint \eqref{Hamiltonian}.

The results corresponding to the six different initial data selections are compared
in Appendix~\ref{app:plots}.

Once we have specified all the initial data, the dynamical trajectories
have been integrated until a final time $t_{\rm fin}$ defined by the relation
$v_\theta(t_{\rm fin})=v_\theta(t_0)$.
Since our primary interest is the behavior of the inhomogeneities,
only the subset of EOM for $\{v_\theta, b_\theta, a_m\}$ has been solved. This is possible
because of the decoupling explained in Subsec.~\ref{sec:eoms}.

We have integrated $400$ dynamical trajectories in case $(i)$ and
$1000$ trajectories in each of the cases $(ii)-(vi)$, thus reaching the total number of
$5400$ trajectories.
In order to test the correctness of the numerical evaluation, we have checked the conservation of
the Hamiltonian and momentum constraints,
as well as of the constants $K_m$ (this check is done by comparing their values
at the initial and final times, $t_0$ and $t_{\rm fin}$).
Imposing this requirement with an absolute precision of $10^{-4}$ turns out to
rule out 831 trajectories, that we consider incorrect, thus leaving us with a final
population of $4569$ dynamical trajectories to analyze.

For each element of this population, we have computed the relative change,
\begin{equation}
  \Delta|a_m|:=\frac{|a_m|(t_{\rm fin})- |a_m|(t_{0})}{|a_m|(t_{0})},
\end{equation}
of the inhomogeneity amplitude $|a_m|$, analyzed separately for each
excited mode. The results have then been compared with the ratio between the actual
bounce point $v_b$ [not to be confused with the solution of Eq.~\eqref{sinbeq}, $v_B$]
and its homogeneous counterpart. This ratio, following from the
analysis of Sec.~\ref{minimum}, provides us with
a measure of how inhomogeneous a given universe is.

\begin{figure*}[tbh!]
  \psfrag{d}{$(|a_m|(t_{\rm fin})- |a_m|(t_{0}))/\sqrt{\hbar}$}
  \psfrag{v}{$\log_{10}(v_b/v_h)$}
  \includegraphics[width=8.7cm]{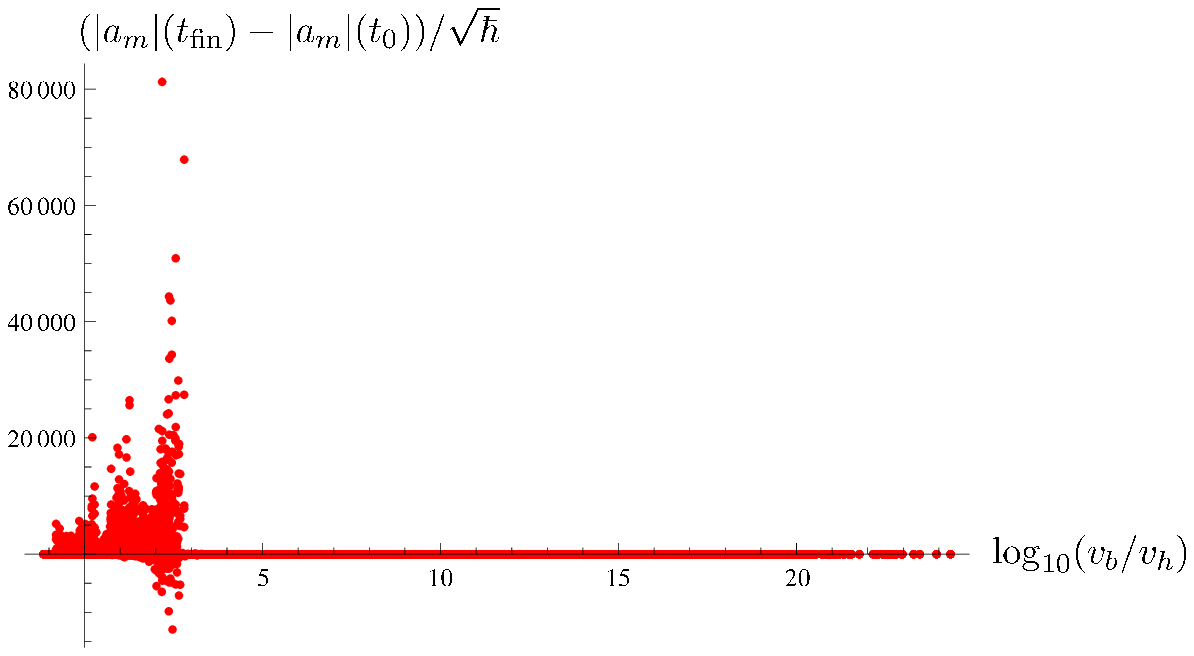}
  \psfrag{d}{$\log_{10}|\Delta |a_m||$}
  \psfrag{v}{$\log_{10}(v_b/v_h)$}
  \includegraphics[width=8.7cm]{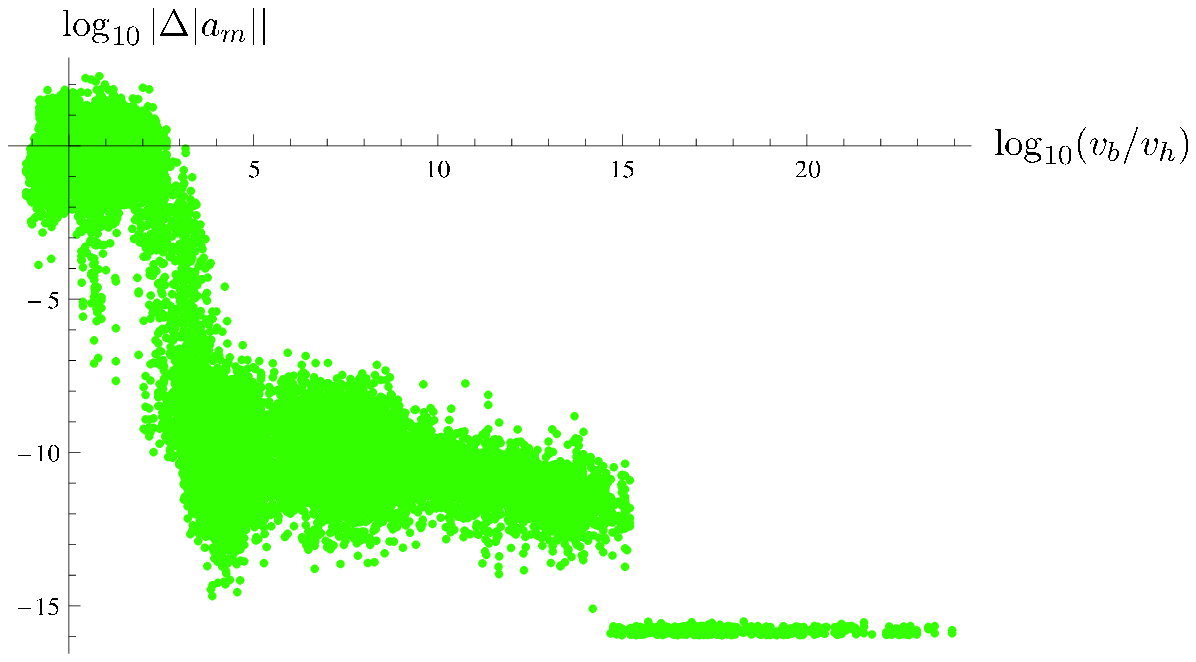}
  \caption{\label{montecarlo} In these plots we show,
    in a linear and a logarithmic scale respectively, the absolute and the relative change
    in the amplitude of the inhomogeneities $|a_m|$ produced when the bounce is crossed, represented against
    the ratio between the actual
    bounce point $v_b$ and its homogeneous counterpart $v_h$. We note that, for
    small $v_b/v_h$ (almost homogeneous universe), $\Delta|a_m|$ can exceed the unity,
    regime which corresponds to the situation when
    the inhomogeneities are amplified. On the other hand, for large $v_b/v_h$ (very inhomogeneous universe),
    $|\Delta|a_m||<10^{-4}$, and thus the amplitude of the inhomogeneity mode is approximately conserved.
    The horizontal line in the right plot corresponds
    to the data points where $|\Delta|a_m||<10^{-16}$, and could not
    be measured accurately owing to floating point precision limits.
  }
\end{figure*}

\begin{figure*}[tbh!]
  \begin{flushleft}
  \psfrag{d}{$\log_{10}|\Delta |a_m||$}
  \psfrag{v}{$\log_{10}(v_b/v_h)$}
  \psfrag{s}{$\sigma=(0,1000)$}
  \includegraphics[width=8.8cm]{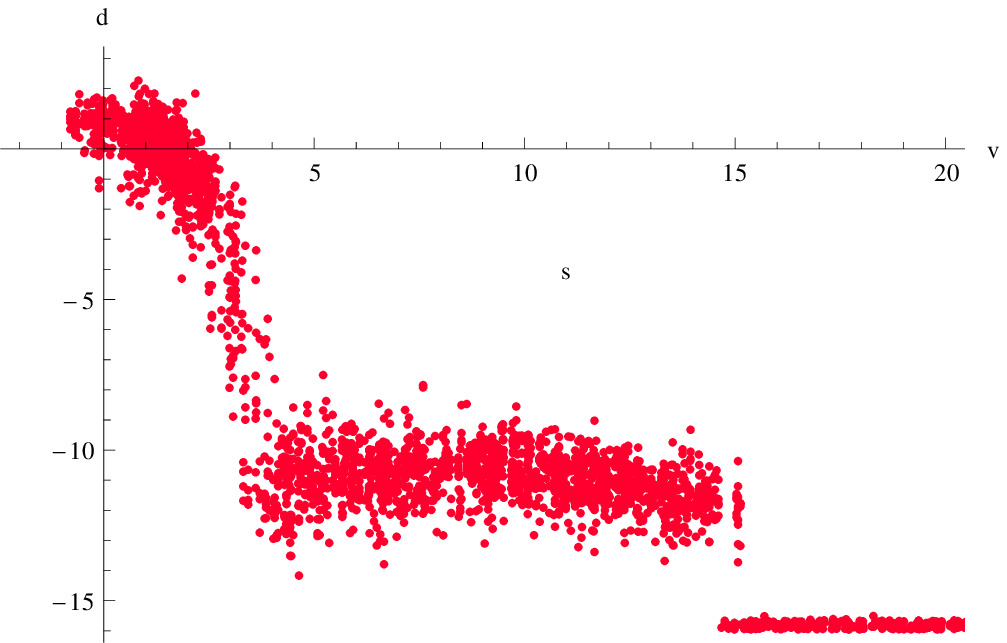}
  \psfrag{d}{$\log_{10}|\Delta |a_m||$}
  \psfrag{v}{$\log_{10}(v_b/v_h)$}
  \psfrag{s}{$\sigma=100$}
  \includegraphics[width=8.8cm]{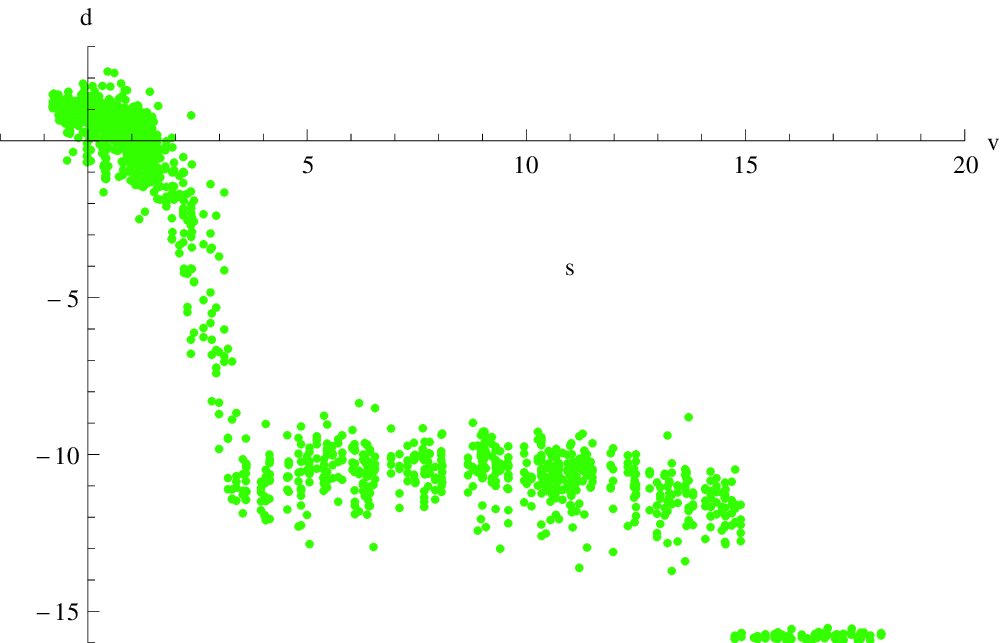}
  \begin{center}\end{center}
  \vspace{-0.6cm}
  \psfrag{d}{$\log_{10}|\Delta |a_m||$}
  \psfrag{v}{$\log_{10}(v_b/v_h)$}
  \psfrag{s}{$\sigma=0.1 v_h^{2/3}$}
  \includegraphics[width=8.8cm]{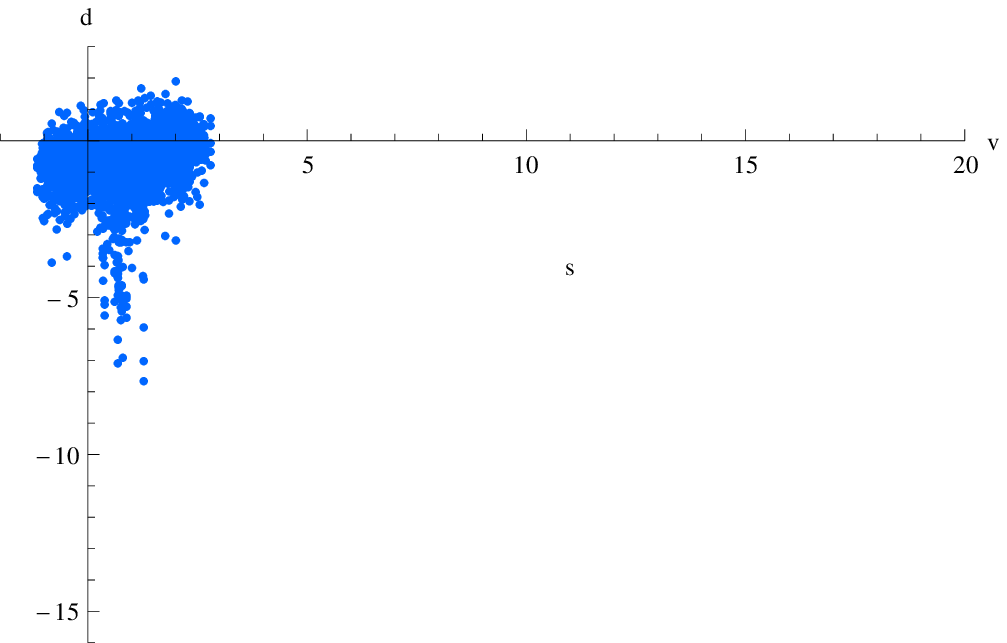}
  \psfrag{d}{$\log_{10}|\Delta |a_m||$}
  \psfrag{v}{$\log_{10}(v_b/v_h)$}
  \psfrag{s}{$\sigma=0.5 v_h^{2/3}$}
  \includegraphics[width=8.8cm]{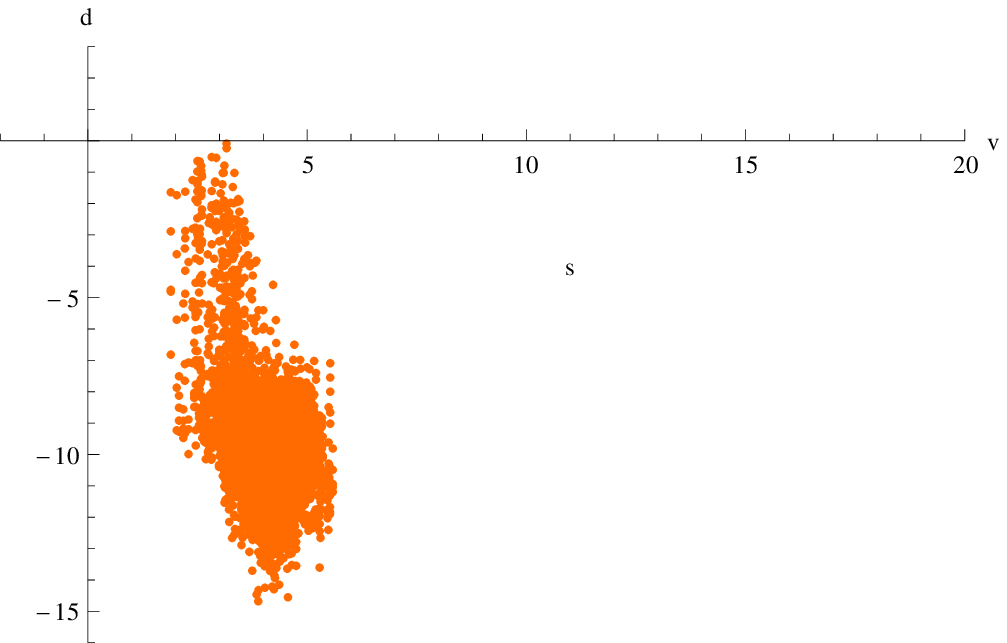}
  \begin{center}\end{center}
  \vspace{-0.6cm}
  \psfrag{d}{$\log_{10}|\Delta |a_m||$}
  \psfrag{v}{$\log_{10}(v_b/v_h)$}
  \psfrag{s}{$\sigma=2 v_h^{2/3}$}
  \includegraphics[width=8.8cm]{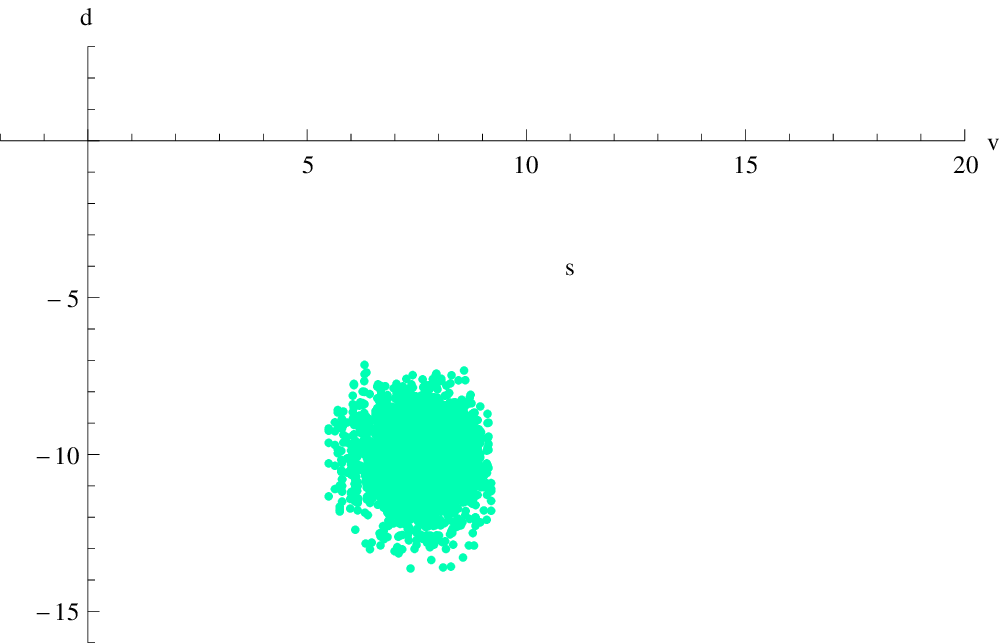}
  \psfrag{d}{$\log_{10}|\Delta |a_m||$}
  \psfrag{v}{$\log_{10}(v_b/v_h)$}
  \psfrag{s}{$\sigma=v_h$}
  \includegraphics[width=8.8cm]{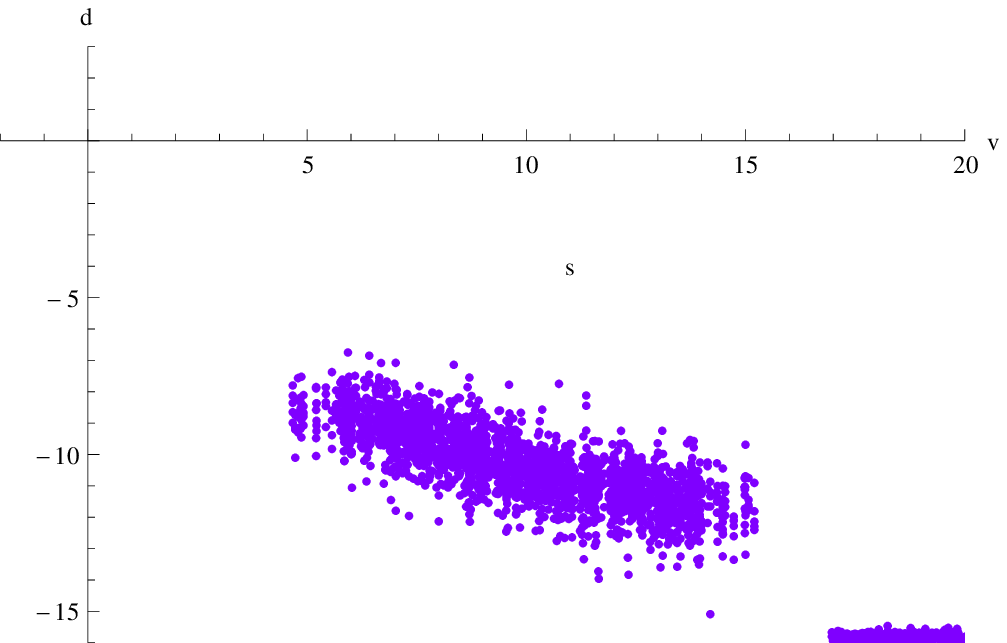}
  \end{flushleft}
  \caption{\label{plots}
    The behavior of the amplification of the inhomogeneities through the bounce
    shown for the population of dynamical trajectories determined by the initial data
    that are generated with each of the algorithms specified in Sec.~\ref{sec:montecarlo},
    listed there as cases $(i)$-$(vi)$.
    Remarkably, for all these cases, the data stay (approximately) concentrated in a specific region
    of the represented plane.
    We can distinguish the near-homogeneous sector (with data concentrated about the origin) and the
    inhomogeneity dominated one (in a patch separated from the former by a sharp drop).
    The cases with $\sigma\propto v^{2/3}$ select a particular sector of that patch. The sharp horizontal line
    about $|\Delta|a_m||=10^{-16}$ has the same origin as in Fig.~\ref{montecarlo}.
  }
\end{figure*}

The final results are displayed in Figs.~\ref{montecarlo} and~\ref{plots}, where $\Delta|a_m|$
is plotted against $v_b/v_h$ for the whole population of trajectories and in each of the
cases $(i)$-$(vi)$, corresponding to the different algorithms used to generate de initial data for the
inhomogeneities.
Looking at Fig.~\ref{montecarlo}, one can distinguish two
distinct regimes:
\begin{enumerate}[(a)]
  \item \emph{inhomogeneity dominated}
    (large $v_b/v_h$), for which the dynamics around the
    bounce is dominated by the content of gravitational waves,
    and \label{it:num-id}
  \item \emph{near-homogeneous} (small $v_b/v_h$), for which
    those waves introduce only small corrections to the
    vacuum Bianchi I dynamics. \label{it:num-nh}
\end{enumerate}
These two regions are clearly separated by an intermediate region where $\Delta|a_m|$ drops sharply.

In the first of these regimes, the relative changes are very small; thus, by the results of
Sec.~\ref{sec:num-gen}, the data points are in the regime where the average of
$\Delta|a_m|$ over the dependence on initial phases is ensured to vanish.
On the other hand, in the second regime above, we see that a large fraction of the data points
exceed $\Delta|a_m|=1$, and hence they correspond to the case where the
antisymmetric behavior of $\Delta|a_m|$ as a function of the phase differences
is generically lost, and its average becomes strictly positive.
Therefore, in the near-homogeneous regime the quantum geometry effects around the bounce
\emph{enhance the inhomogeneities}.

An intuitive explanation of this mechanism comes from the analysis of the
previous subsection. As far as we
have been able to observe, the maximum change that a given mode $|a_m|$
can experience at the bounce is of the order of the amplitude of
its superimposed ``oscillations'' in the nearby region.
This latter amplitude increases like $v_\theta^{-\epsilon}$, and
nearly homogeneous universes have much smaller value of $v_\theta$ at
the bounce than the highly inhomogeneous spacetimes. This allows for larger relative
changes of $|a_m|$. When the amplitude of the oscillations in Fig.~\ref{dephase} exceeds
the mean value of $|a_m|$, the antisymmetry in the dependence on the phase difference is lost,
because $|a_m|$ must always remain positive by definition. The breaking of this
antisymmetry opens the possibility of an amplification.

Let us finally remark that the minimum bounce point that we have observed in the studied population of
dynamical trajectories is
$v_{b}=0.070 v_h$. This value is consistent with the lower bound given
in Sec.~\ref{minimum} for the bounce point and, in fact, it is quite close to it.

\section{Summary and discussion}\label{sec:concl}

We have studied a class of cosmological inhomogeneous spacetimes --the vacuum Gowdy universes
of $3$-torus topology-- taking as
starting point a first order
effective theory, built out of a quantum model that has been obtained by implementing
the hybrid quantization scheme of LQC.
It is worth emphasizing that the presented treatment constitutes the first nonperturbative effective
description of an inhomogeneous spacetime within the loop framework which has been constructed in a systematic and
controllable way out of a rigorously defined quantum system.
Using this description, we have investigated the dynamics of a large population
of universes, focusing on two issues.

The first of them is the verification of the persistence of the bounce phenomenon observed in homogeneous
LQC in spite of the presence of inhomogeneities. With that aim, we have compared the properties of
an inhomogeneous universe with those of its Bianchi I counterpart, namely, the homogeneous background on
which the gravitational waves responsible for the inhomogeneities propagate.
The analysis of the EOM and the
constraints have proven that the bounce is indeed present, perhaps with the possible exception of certain critical
trajectories.
In two of the directions in configuration space (defined in terms of the areas of large
$2$-tori) the bounce of the Gowdy universe and its Bianchi I counterpart coincide.
For the remaining
(``inhomogeneous") direction, the exact results depend on the considered region of phase space. In
particular, in the region where the constants of motion $\Theta_\delta$ and $\Theta_\sigma$
[see Eq. \eqref{eq:Theta-eff}] satisfy the inequality $\Theta_\delta\Theta_\sigma<0$,
the bounce always happens further away from the analog of the classical singularity
than in the Bianchi I counterpart.
In the other case, $\Theta_\delta\Theta_\sigma>0$, the bounce can actually happen closer to the singularity,
and determining its properties has required a combined analytical and numerical analysis. We have shown then
that there exists a lower bound on the ratio between the bounce positions in our system and in its
Bianchi I counterpart. In terms of the parameters $v_\theta$ \eqref{eq:new-vb}, this bound
is approximately $0.05$, and is valid in the whole space of solutions,
except possibly for an extremely small region around the zero measure set of critical
trajectories.

Let us now comment on those critical solutions. They represent a very special and
quite peculiar class of trajectories, as the universe following them actually can
collapse all the way down to a singularity through a very intricate,
infinite sequence of bounces and recollapses. However, owing to the intrinsic limitations
of the effective treatment, they cannot be considered as physical.
Indeed, they pass through the bifurcation points, where minute changes of initial data can modify the global trajectory
drastically. In consequence, the effects of the higher-order state dependent parameters, here neglected in the effective
description, can play a significant role in theses cases,
and the effective description itself ceases to be valid.

Once the presence of the bounce as a generic feature has been positively confirmed in the considered model,
we have addressed the second issue, namely, whether there exist significant changes to the structure
of the inhomogeneities across the bounce. Here, our
investigation has been mainly focused on comparing the distributions of the amplitudes of the inhomogeneity modes
(which encode the information about the distribution of the energy of the gravitational waves)
in the distant future and past of the universe.
To discuss this question, we have converted the deterministic dynamical system into a statistical one
by averaging over phases in each of the modes. In order to study the properties of this statistical system,
the Monte-Carlo method has been applied. First, a large population of dynamical
trajectories has been calculated numerically.
And then, the data representing the asymptotic behavior of the inhomogeneity modes has
been extracted and used to probe a statistical relation between the relative change
of the amplitude of the modes and the measure of the departure from homogeneity.

Our analysis has revealed the existence of two separate regions in the space of solutions.
One of them, denoted as inhomogeneity-dominated, contains those Gowdy universes which bounce
at volumes much greater than their homogeneous counterparts, and for which the bounce process
itself is driven by the inhomogeneities. In that regime, the amplitude of the modes
is statistically preserved. On the other hand, in the near-homogeneous region, where the
inhomogeneities produce only small corrections to the evolution of the homogeneous background,
the inhomogeneities turn out to be statistically amplified.
The mechanism of this amplification has been understood
by analyzing the behavior of individual modes.
Remarkably, in our statistical investigation, we have not detected any measurable difference
that could be assigned to a dependence on the mode number of the inhomogeneity.

The above results provide a potentially interesting mechanism for structure generation.
If the universe is sufficiently homogeneous, the
bounce process enforces the growth of inhomogeneities, in this case pumping energy
to the gravitational waves. Only when those inhomogeneities reach a sufficiently large amplitude,
the growth stops and their structure stabilizes.

The very same mechanism allows one also to shed some light on the process of
entropy growth through the bounce. Indeed, in the system
under study, the energy of the inhomogeneities provides an intuitive entropy measure.
The results of the above paragraph can be thus
reinterpreted in the following way. In the process of the bounce, the entropy of the universe would increase
till it reaches certain critical order of magnitude. Then, the bounce process would preserve it.
Of course, this statistical mechanism is applicable only for populations of universes,
not for a single one. However, it may become feasible in the case of classically
recollapsing models, like for example Gowdy universes of the
topology $S^1\times S^2$ and $S^3$ \cite{cmv-GS-uniq,*bgv-G-unitary}. In these cases,
the infinite chain of bounces and recollapses (within
the evolution of one universe) can make the statistical process physically meaningful.

All these results, though very promising, should be treated only as preliminary
for various reasons, that we now comment.

First, our effective treatment is only a first-order theory,
where all the state-dependent quantities have simply been
dropped. Therefore, its applicability and reliability needs to be further confirmed,
either by a systematic derivation of the effective dynamics with careful estimates
of higher-order corrections or by a comparison with the genuine quantum evolution
in a domain sufficiently large as to be representative. We already know from our
discussion in this section that the treatment is to fail in the near-critical
regime. The verification becomes particularly important, since we are considering
a system with an infinite number of degrees of freedom, and it is
only in a series of systems with a finite number of them that
it has been possible to verify so far the correctness of the effective dynamics.
Fortunately, in the considered model,
the structure of the dynamics of the inhomogeneous degrees of freedom is sufficiently
simple (modes couple only in pairs) to believe that it is possible to extrapolate the
results about the validity of the effective description from the quantum
mechanical systems. Furthermore, the decoupling between (pairs of) modes opens the
possibility of verifying this validity by applying technically manageable inductive methods,
based for example on a combined effective/genuine treatment\footnote{One starts with
the genuine analysis of just one pair of modes. Then, the complexity of the system is
increased in an iterative process where new modes are added, keeping the total number finite,
and in which all but the pair of highest mode number are controlled by an effective theory
(applying the results of the previous step in the construction), whereas that last pair
remains fully quantum.}, or adapting the systematic effective
treatment of Ref. \cite{bss-eff1,*bt-eff2,*bbhkm-eff3}.
This will be the subject of future investigations.

The above limitations are connected with an important
and potentially relevant physical problem: the zero point energy.
Namely, as discussed in Sec.~\ref{sec:frame-hybrid}, the inhomogeneity
modes form a Fock space and each of them (after having implemented a
convenient conformal transformation) behaves like a
scalar field with a time dependent mass,
with a ground state that is not dynamically invariant.
As a consequence, one may expect the (dynamical) generation of a nonzero energy
density, which could radically affect the system
once all the modes are taken into account. This effect,
certainly happening in some geometrodynamical systems
when one adopts certain choices for the Fock quantization
of the inhomogeneities, is currently
under investigation. Concerning the choice of Fock quantization,
it is worth emphasizing again that the quantization adopted here is
the only one that ensures a unitary quantum
evolution --without (at least some) divergences-- after a suitable
deparametrization, which results from a convenient choice
of internal time. On the other hand, introducing a polymer quantization
for both the geometry and the matter fields (see e.g.
\cite{klb-scalar1,*klo-scalar2,*hhs-scalar} for the latter) may even
be important in order to achieve a satisfactory regularization of
any matter field divergent contribution.
However, along the lines that we have commented, the level
or even the existence of the nonzero energy ground state seems to
strongly depend on the choice of an evolution parameter and of
the partial observables. This effect may thus restrict possible choices
within the genuine quantum theory, ruling out some selections of
an internal time as leading to unphysical results \cite{p-prep}.

Another issue which deserves some comments is the quantization of the
homogeneous degrees of freedom. As we have discussed, in the process of
quantizing the Bianchi I background \`a la loop, we have applied
the old quantization prescription for the improved dynamics, proposed
in Refs. \cite{ch-b1,chv-b1}. Here, the compactness of the system protects
us against any physical inconsistency regarding the dependence on the
choice of a fiducial cell in the construction. However, it would seem
unnatural to expect that the compact topology by itself may justify using
a prescription that is invalid in the noncompact cases. Fortunately, the
scheme valid for the noncompact systems \cite{awe-b1} is sufficiently similar
in the aspects that are relevant for our analysis, and hence the qualitative results
reported here must hold in the model built with the new scheme as well \cite{mmw}.
Nonetheless, the quantitative results, like the exact bound on the bounce, the scale
at which the inhomogeneities reach the equilibrium, or the critical behavior, are
expected to change. Therefore, these qualitative results should be viewed
as tentative only. The main aim of the article is to provide and test the methodology,
which will be applied in more detailed studies of the improved system, exploring and
confirming the kind of phenomena found here.

The vacuum Gowdy universes, while simple and useful to test the formalism and develop
the methodology, are not fully physical from the viewpoint of observational cosmology,
since they do not admit near isotropic solutions. Furthermore, the degeneracy
of the inhomogeneity modes qualitatively differs from the observed one, which has a
completely distinct structure. The first problem is solved with the inclusion of matter.
For that purpose, coupling a massless scalar field to the Gowdy universes
is a particularly appealing possibility. Building an effective theory out
of (the already found \cite{mmm}) hybrid quantum description of this Gowdy model with matter
and repeating the dynamical analysis performed here will be the next step in our
investigations. The second problem, however, requires the study of models
where the energy level degeneracy is that of the spherical harmonics.
For that, one has to go beyond the nonvacuum Gowdy models, considering other more complicated
families of spacetimes containing matter fields and where the inhomogeneities are not restricted
by the existence of isometries.

\begin{acknowledgments}
  The authors are grateful to A.~Ashtekar, J.~Cortez, L.J.~Garay,
  M.~Mart\'in-Benito, and J.M.~Mart\'in-Garc\'ia
  for discussions.
  D.B. acknowledges financial support from the Spanish
  Ministry of Education through the \emph{Programa Nacional de Movilidad de
  Recursos Humanos} of National Programme No. I-D+i2008-2011.
  This work was supported by the MICINN Project
  FIS2008-06078-C03-03 and the Consolider-Ingenio Program
  CPAN (CSD2007-00042) from Spain, by the Institute for Gravitation and the Cosmos (PSU),
  and by the Natural Sciences and Engineering Research Council of Canada.
\end{acknowledgments}

\appendix

\section{Different types of data for the Monte-Carlo simulations}
\label{app:plots}

In this appendix we explain what kind of universes one obtains
starting with the different types of initial data
of Sec.~\ref{sec:montecarlo}. More specifically, we characterize the universes with
the ratio $v_b/v_h$, which is small (large) for near-homogeneous
(inhomogeneities-dominated) universes.
As can be observed in the first two
plots of Fig.~\ref{plots}, when we choose $\sigma$ independent of
$v_h$ we get the whole range of possible cases, containing both
near homogeneous and highly inhomogeneous universes.
On the other hand, when we
choose $\sigma= a v_h^{2/3}$, our procedure selects only
a particular region (depending on $a$) which contains just
universes peaked around a specific ratio $v_b/v_h$.
The choice $\sigma\propto v_h$, on the other hand, leads again
to solutions with a broad range for $v_b/v_h$, but now strongly inhomogeneous
universes are generated, owing to the fact that $\sigma$ is generically large.
Independently of the method of generation, the data seem to stay in the same
region: a well defined patch in the plane $\Delta|a_m|$--$(v_b/v_h)$.

\bibliography{bmp-GowdyEff}{}
\bibliographystyle{apsrev4-1}

\end{document}